\documentclass[10pt,a4paper]{article}
\usepackage{graphicx}
\usepackage{amssymb}
\usepackage{slashed}
\usepackage{multirow}
\usepackage{hhline}
\usepackage{makecell}
\usepackage{comment}

\usepackage{caption}

\newcommand{\be}{\begin{eqnarray}}
\newcommand{\ee}{\end{eqnarray}}
\usepackage[colorlinks]{hyperref}
\usepackage{cancel,bm}

\newcommand{\sch}{Schr\"odinger}
\usepackage{amsmath,amssymb}
\usepackage{mathtools}
\usepackage{float}
\usepackage{color}
\usepackage{leftidx}
\usepackage{booktabs}
\usepackage{latexsym}
\usepackage{multirow}
\usepackage{hypcap}
\usepackage{array}
\usepackage{xfrac}
\usepackage{soul}
\usepackage{cite}
\usepackage[title]{appendix}
\usepackage{authblk}
\hypersetup{
    colorlinks=red,
    citecolor=blue,
    filecolor=magenta,
    urlcolor=blue,
}

\usepackage[a4paper,top=2cm,bottom=2cm,left=2cm,right=2cm]{geometry}

\begin{document}
\title{The pion in the graviton soft-wall model: 
phenomenological applications }

\author[1]{Matteo Rinaldi 
 \footnote{Corresponding author email: matteo.rinaldi@pg.infn.it} }

 \author[2]{{ F. A. Ceccopieri}}
\author[3]{Vicente Vento}

    \affil[1]{  \rule[25pt]{0pt}{0pt} Dipartimento di Fisica e Geologia. Università degli studi di
 Perugia. INFN section of Perugia. Via A. Pascoli, Perugia, 06123, Italy.
}

        \affil[2]{ {  \rule[25pt]{0pt}{0pt} Universit\'e Paris-Saclay, CNRS, IJCLab, 91405, Orsay, France \\ IFPA. Univèrsité de Liège, B4000, Liége, Belgium.}}

\affil[3]{ \rule[25pt]{0pt}{0pt} Departamento de F\'{\i}sica Te\'orica-IFIC, Universidad de Valencia- CSIC,
46100 Burjassot, Valencia, Spain. }

\date{}

 \maketitle

\begin{abstract}

 The holographic graviton soft-wall model, introduced to
describe the spectrum of scalar and tensor glueballs, is improved
to incorporate the realization of chiral-symmetry as in QCD. 
Such a goal  is achieved by including 
 the longitudinal dynamics of QCD into the scheme. Using the relation between AdS/QCD and Light-Front 
dynamics, we construct the appropriate wave function for the pion which is used to calculate several pion observables. The 
comparison of our results with phenomenology is remarkably successful.

\end{abstract}

\section{Introduction}

In the last few years, hadronic models, inspired by  the holographic conjecture 
\cite{Maldacena:1997re,Witten:1998zw}, have been vastly used and developed in order 
 to investigate 
non-perturbative features of glueballs and mesons, thus trying to 
grasp fundamental features  of QCD 
\cite{Fritzsch:1973pi,Fritzsch:1975wn}. Recently we have 
used the so 
called 
AdS/QCD models to study the meson and glueball spectrum
spectrum~\cite{Rinaldi:2017wdn,Rinaldi:2018yhf,Rinaldi:2020ssz,Rinaldi:2021dxh}. The holographic principle relies in a correspondence between a 
five dimensional 
classical theory with an AdS metric and a supersymmetric
 conformal quantum 
field theory with $N_C  \rightarrow \infty$. This theory, different from QCD, is taken as a starting point  to construct a 5 dimensional 
holographic dual of it. This is the so called bottom-up approach~
\cite{Polchinski:2000uf,Brodsky:2003px,DaRold:2005mxj,Karch:2006pv,Erlich:2005qh}. 
The relation of this approach established with QCD is at the level of the 
leading order in the number of colors expansion.

In our previous investigation, we could  successfully describe the pseudo-scalar spectrum, identified with the $\eta$ 
system, without any free parameter~\cite{Rinaldi:2021dxh} within the GSW model. However, we found that the conventional model was not able to distinguish between the spectrum of 
the $\eta$ and the $\pi$~\cite{Rinaldi:2021dxh}.
One of the 
  differences between the $\eta$ and the $\pi$ is the isospin, however since the  GSW does 
not take into account Coulomb corrections, the pions behave very much like the $\eta$ from the point of view of quantum numbers and therefore the 
spectrum  would be the same, but certainly not in nature.  In order to characterize the pion we proposed at that time a modification of the dilaton  in analogy with previous investigations~\cite{Erlich:2005qh,Gherghetta:2009ac,Vega:2016gip}. Such a procedure
was able to describe correctly the pion spectrum, { however we found out that the wave function derivable from the mode function was not precise enough to explain many of the data that follow.  Other authors have also studied chiral symmetry breaking within holographic models in the Hard Wall approach in an aim to match the high and the low energy behavior or QCD~\cite{Sakai:2004cn,DaRold:2005mxj,Hirn:2005nr,Hirn:2005ub}.

In here we will proceed in a way which follows closer QCD by matching our AdS/QCD model with Light Front holography.}
The pion is  the Goldstone boson of SU(2) x SU(2) chiral symmetry and this fact is instrumental in giving the lightest 
pion its low mass.  We should therefore implement the spontaneous broken realization of chiral symmetry in the  GSW 
holographic model to reproduce the low mass of the ground state  of the pion.  To do so we will modify the dilaton of  the GSW model  
to get a zero mass pion. To implement chiral symmetry breaking we will include the QCD longitudinal 
dynamics~\cite{tHooft:1974pnl,Li:2015zda,Burkardt:1997de,Li:2021jqb,deTeramond:2021yyi}. 
In this way we will obtain a pion at the physical mass. We also obtain the corresponding mode function.

In order to establish to what extent  the mechanism is phenomenologically successful, we calculate several pion observables
comparing the outcomes with data. To
this aim,  a correspondence between the mode function and a light cone wave functions is introduced~\cite{Brodsky:1997de,Brodsky:2006uqa,Brodsky:2007hb}.
The comparison between our calculations and the phenomenological results is quite reasonable despite the low number of parameters.

The contents of this presentation go as follows. In  Section \ref{eta} we describe the $\eta$ Equation of Motion (EoM) of the GSW model and its solution. We proceed in Section \ref{pion} to show the description of the Goldstone pion in the GSW model. In Section \ref{pionLF}
 we establish the relation between the pion mode function and the pion light cone wave function (wf).
In order to effectively describe the
  chiral symmetry breaking, the longitudinal dynamics is introduced in Section \ref{longitudinal}. Having completed the description of the pion model, in  Section \ref{observables} we calculate various pion observables: the spectrum, the form factor, the  mean square radius, the effective form factor, the mean transverse distance 
between  two partons,
 the decay constant, the distribution amplitude, the  photon transition form factor  and the parton distribution function (PDF). Results are compared with the phenomenology. We end by some concluding remarks.

\section{The pseudo-scalar wave equation within the GSW model}
\label{eta}
In this section, details on the application of the GSW model  to the study of the pseudo-scalar meson spectrum are presented. 
Our approach is based on the usual Soft Wall (SW) AdS/QCD model
 \cite{Brodsky:2007hb,Karch:2006pv} modified by  a deformation of the $AdS_5$ space.
These type of changes have been proposed in several analyses 
to improve the prediction power of the holographic models 
within the bottom-up framework~
\cite{Andreev:2006vy,Capossoli:2015ywa,FolcoCapossoli:2019imm,MartinContreras:2021yfz,Ghoroku:2005vt}.
 In particular, the GSW model
has been specifically introduced to describe the scalar and tensor 
glueball spectrum \cite{Rinaldi:2017wdn}. Previously it was shown that the conventional field approach was hardly capable to describe the glueball 
spectra~\cite{Capossoli:2015ywa,Colangelo:2007pt}. Therefore, in Ref. \cite{Rinaldi:2017wdn} the glueball masses 
have been calculated from the mode function of gravitons propagating
on a deformed  $AdS_5$ space. The wrap metric of the model effectively encodes complex dilatonic effects. In particular the metric used was

\begin{equation}
ds^2=\frac{R^2}{z^2} e^{\alpha k^2 z^2} (dz^2 + \eta_{\mu \nu} 
dx^\mu dx^\nu) =  e^{\alpha {k^2 z^2}} g_{M N} dx^M 
dx^N  =\bar{g}_{MN}dx^M dx^N.
\label{metric5}
\end{equation}
Recently,  in Refs. \cite{Rinaldi:2018yhf,Rinaldi:2020ssz,Rinaldi:2021dxh} the GSW has been applied to 
describe also the spectrum of various  mesons and high spin glueballs. The parameters entering the GSW model
are $\alpha = 0.55 \pm 0.04$  and the energy scale $k=0.37/\sqrt{\alpha} $ GeV.
Within these values
 the model is able to reproduce 
quite accurately the meson and glueball spectra, except for the pion   which requires proper modifications to include the chiral symmetry breaking mechanism~\cite{Rinaldi:2021dxh}.
In this case however, the pion spectrum comes out close to the experimental 
data although the calculation predicts the existence of up to now not found additional states.

Let us proceed now to build up the model for the pion, which is closer to the phenomenology than that of Ref. \cite{Rinaldi:2021dxh}.
We start from the pseudo-scalar action,

\begin{align}
\label{Eq:action}
 S = \int d^5x~ e^{-\varphi_0(z)} \sqrt{-g} \Big[g^{MN} \partial_M \Phi(x)
\partial_N \Phi(x)-4 e^{\alpha k^2 z^2} \Phi(x)^2 \Big]~,
\end{align}
where $\varphi_0(z)=k^2 z^2$ and we have explicitly used the 
 five dimensional pseudo-scalar
mass $M_5 R^2 = -4$~\cite{Contreras:2018hbi}.  The EoM for the pseudo-scalar system can be recast in a 
Schr\"odinger type equation by scaling the field,

\begin{align}
\label{Eq:TR1}
      \Phi(x,z)=  e^{iP \cdot x} e^{\varphi_0(z)/2} \chi(z)z^{3/2} ,
\end{align}
where $P^2= M^2$, $M$ being  the mass of the pseudo-scalar meson.
The final equation has the form

\begin{equation}
-\frac{d^2 \chi(z)}{dz^2} +V(z)  \chi(z) = M^2 \chi(z).
\label{psmeq0}
\end{equation}
where the potential is

\begin{equation}
V(z) =  k^4z^2 + 2 k^2 + \frac{15}{4z^2} 
- \frac{4}{z^2} e^{\alpha k^2 z^2} .
\label{etapot}
\end{equation}

However,
since the potential Eq.(\ref{etapot})  is not binding, 
an additional dilaton  $\varphi_n(z)$ has been added to the action
 Eq. (\ref{Eq:action}) 
so that the exponential $exp[\alpha k^2 z^2]$ in Eq. (\ref{etapot}) 
 can be truncated  so that the  final potential   binds. 
Details on this procedure have been described in detail in Ref. \cite{Rinaldi:2021dxh}.
The \sch~equation is now obtained by imposing the following re-scale:

\begin{align}
\label{Eq:TR2}
      \Phi(x,z)=  e^{iP \cdot x} e^{(\varphi_0(z)+\varphi_n(z))/2}~ \chi(z)z^{3/2} ,
\end{align}
then,  the truncated phenomenological potential becomes:

\begin{equation}
V(z) =  k^4z^2 (1-2 \alpha^2) + 2 k^2 (1-2\alpha) - \frac{1}{4z^2} 
\label{psmeq}
\end{equation}
The regular solution to the above differential equation, for the $n$th mode  is,

\begin{align}
 \chi_n(z)
&=N_n z^{ \frac{1}{2} } e^{-\frac{\sqrt{1-2 \alpha^2}}{2}k^2 z^2  } ~ _1F_1 \left(-n,1, 
\sqrt{1-2 \alpha^2} k^2 z^2
 \right)~,
 \label{etawf}
\end{align}
where $N_n$ is a normalization constant and the pseudo-scalar mass comes out,

\begin{align}
 M(n)^2 =  \Big[(2-4 \alpha)+2\sqrt{1-2\alpha^2}(1+2 n)\Big]k^2.
\end{align}
It is immediate that, $M(0) \ne 0$, a consequence of the fact that the model does not 
incorporate chiral symmetry. We identify this state with the dual of
$\eta$ meson. The spectrum, shown in Ref. \cite{Rinaldi:2021dxh} is in good agreement with data.

\section{The pion in the GSW model}
\label{pion}
In QCD if the quark masses are zero, chiral symmetry is spontaneously broken and the pion is the corresponding Goldstone boson. Therefore if we want an AdS/QCD 
model which represents this QCD behaviour we have to obtain an EoM for a massless pion.
In this holographic framework
the dilaton must describe much
the essence of the
  confinement mechanism together with the  chiral symmetry 
breaking. Therefore,  our formalism, connecting the pseudo-scalar spectrum to that of the $\eta$,
requires a modification of the dilaton to generate the exact chiral limit.
From a phenomenological point of  view, it is necessary to get a massless pion and an energy scale bigger
then that of the $\eta$, as signaled by pion mass hierarchy. For this
purpose, we propose a straightforward modification of the
additional dilaton $\varphi_n(z)$ used for the $\eta$~\cite{Rinaldi:2021dxh}. It will be sufficient  to introduce two 
additional constants in the cut off potential.
Therefore, for the pion within the GSW model we propose the following action:

\begin{align}
\label{Eq:action2}
 S = \int d^5x~ e^{-\varphi_0(z)-\varphi_n(z)} \sqrt{-g} \Big[g^{MN} \partial_M \Phi(x)
\partial_N \Phi(x)-4 e^{\alpha k^2 z^2} \Phi(x)^2 \Big].
\end{align}
If   the additional dilaton satisfies the following
differential equation (see Ref. \cite{Rinaldi:2021dxh} for details):

\begin{align}
-\frac{\varphi_n^{''}(z)}{2} +\varphi_n^{'}(z) \left( \frac{3}{2z} 
+ k^2 z  \right)  + \frac{\varphi_n^{'}(z)^2}{4} 
- \frac{4}{z^2} \left[ e^{\alpha k^2 z^2}-1-(\alpha + \xi_\pi)
k^2 z^2- \frac{1}{2}( \alpha^2 + \gamma_\pi) k^4 z^4  \right]=0~,
\label{difeqs1}
\end{align}
where the parameters $\xi_\pi$ and $\gamma_\pi$ have been included
at variance of the $\eta$ case  \cite{Rinaldi:2021dxh}.
The relative potential in the Schr\"odinger representation will be:

\begin{align}
V_\pi(z) = \frac{15}{4 z^2} +2 k^2+k^4 z^2 - \frac{4}{z^2}\left[1+(\alpha 
+ \xi_\pi)k^2 z^2+\frac{1}{2}( \alpha^2 + \gamma_\pi) k^4 z^4 \right] 
= V_\eta(z)-4 k^2 \xi_\pi-2 \gamma_\pi k^4 z^2~.
\label{pionpotential}
\end{align}
For this potential the mass equation becomes

\begin{equation}
M_\pi^2(n)= \big[2- 4(\alpha +\xi_\pi) + 2 \sqrt{1- 2(\alpha^2 + \gamma_\pi)}(1+2n)\big]k^2.
\end{equation}
If  one imposes $M_\pi(0) =0$ then:
\begin{align}
 \xi_\pi= \dfrac{1-2 \alpha+\sqrt{1-2 \alpha^2-2 \gamma_\pi}}{2}.
\end{align}
This relation ensures that the lightest pion is a Goldstone boson.
The mass spectrum then becomes
\begin{align}
 M^2_\pi(n) = 4\sqrt{1-2 (\alpha^2+ \gamma_\pi)}~k^2n ~.
\label{mpi} 
\end{align}
Thus, the parameter $\gamma_\pi$ modifies the $\eta$ slope of the spectrum to reproduce the pion 
excitations. This freedom relaxes the energy scale from the GSW scale $\sqrt{\alpha}k$.
The $n$th solution to the relative \sch ~equation is:

\begin{align}
 \chi_n(z)
&=N_n z^{ \frac{1}{2} }~ exp\left(-\frac{\sqrt{(1-2 (\alpha^2- \gamma_\pi)}}{2}k^2 z^2  \right) ~ _1F_1 \left(-n,1, 
\sqrt{(1-2 (\alpha^2+\gamma_\pi)} k^2 z^2
 \right)~.
 \label{pionwf}
\end{align}
where the difference with respect to Eq.(\ref{etawf}) is characterized by the presence of $\gamma_\pi$.

A caveat is in place. The experimental status for the pion excitations is not well established and therefore possible intermediate states 
between $\pi$, $\pi'$ and $\pi''$  could be observed in the future as discussed in Ref. \cite{Rinaldi:2021dxh}.
However, if one tries to describe the present experimental spectrum as described in Ref.  \cite{Tanabashi:2018oca,Zyla:2020zbs}, 
different energy scales for the scalar, the $\eta$ and the pion spectra 
are necessary. The modification proposed here preserves entirely the GSW 
description of the pseudo-scalar structure but incorporates a new scale and
a massless pion ground state. In the future other different possibilities will be investigated. In the present study we 
mainly focus on this strategy which preserves as much as possible
the GSW structure of the model. In the next sections  phenomenological predictions and comparisons with observable will be provided.

\section{Pseudo-scalar Light Front wave function} 
\label{pionLF}
In order to test the proposed approach,  let us  take advantage of the correspondence between the AdS/QCD approach and the Light-Front 
formalism that characterizes the non perturbative 
structure of hadrons~\cite{Brodsky:1997de,Diehl:2000xz}.
Such a strategy is fundamental to use the  AdS/QCD model
to evaluate other observables and  learn new information 
on the inner structure of the hadrons.
Here we recall how  to translate the mode function derived from the AdS EoM, e.g. from Eq. 
(\ref{pionwf}) for the pion, in terms of the 
corresponding Light-Front 
(LF) wave function  \cite{Brodsky:2006uqa,Brodsky:2008pf,deTeramond:2008ht}. This procedure is extremely convenient for 
calculating observables  to test the proposed models.
 In particular, we 
 follow the formalism presented in Ref. \cite{Brodsky:2007hb} where such a 
procedure has been applied to the hard-wall (HW) and 
 SW models in order to describe the pion and the
bulk-to-boundary propagator (the dual to the electromagnetic 
conserved current). 

\subsection{The light front formalism}

Let us first recall the main essence of the LF Fock representation of hadronic 
systems. As shown in Ref. ~\cite{Brodsky:1997de} the QCD quantization at fixed LF 
time $\tau =t+x_3/c$ allows to describe the hadron spectrum from a 
Lorentz-invariant hamiltonian: $\hat H_{LFQCD}= P^- P^+-\mathbf{P}_\perp^2  $,
where the hadron four momentum, described with LF coordinates, has been 
introduced: $P^\mu=(P^+,P^-,\mathbf{P}_\perp)$ and $P^\pm = P_0 \pm P^3$. The plus and transverse components are kinematical operators. 
 $P^- = i d/d\tau$ is responsible for the LF time evolution of the system. The mass equation can be then introduced as

\begin{align}
 \hat H_{LFQCD} |\psi_h \rangle = M^2_h |\psi_h\rangle~,
\end{align}
where $|\psi_h\rangle$ represents the hadron state. Remarkably,  the LF 
quantization, if the $A^+=0$ gauge is considered, leads to 
a suppression of  all intermediate gluon 
degrees of freedom justifying a Fock expansion
of the hadron state in terms of free partons. Therefore:

\begin{align}
 |\psi_h  (P^+, \mathbf{P}_\perp, S_z)\rangle &= \sum_{n, \lambda_i}
\prod_{i=1}^n
\int \dfrac{dx_i d^2 \mathbf{k}_{\perp i} }{2 \sqrt{x_i}(2 \pi)^3}(16 \pi)^3
\delta\left(1-\sum_{j=1}^n x_j \right) \delta^{(2)} \left(\sum_{j=1}^n 
\mathbf{k}_{\perp i}  
 \right)
\\
\nonumber
&=
 \psi_{n/h}(x_i, \mathbf{k}_{\perp i},\lambda_i) 
|x_i P^+, x_i 
\mathbf{{P}_\perp}+ \mathbf{k}_{\perp i}, \lambda_i \rangle_n;
\end{align}
where here $\mathbf{k}_{\perp i}$ is the intrinsic transverse momentum of the
$i$ parton, $\lambda_i$ is the helicity, $n$ the number of Fock states taken in the sum, e.g. for mesons $n \ge 2$.  $\psi_{n/h}$ is the
frame independent LF wf which incorporates the probability that the hadronic
system can be described by $n$ constituents.
$|k^+_i, \mathbf{K}_{\perp i}, \lambda_i \rangle_n $ represents the Fock
state of $n$ free partons, which is an eigenstate of the free LF Hamiltonian. 
The normalization condition reads,

\begin{align}
 \langle \psi_h  (P^+, \mathbf{P}_\perp, S_z)|
\psi_h  (P'^+, \mathbf{P'}_\perp, S'_z)\rangle = 2 P^+ (2 \pi)^3
\delta_{S_z~ S'_z} \delta(P^+-P'^+) \delta^{(2)}(\mathbf{P}_\perp -
\mathbf{P'}_\perp ),
\end{align}
and leads to,

\begin{align}
 \sum_{n=2}^\infty \prod_{i=1}^n \int  \dfrac{dx_i \mathbf{k}_{\perp i}
  }{2 (2 \pi)^3} (16 \pi^3)  \delta\left(1-\sum_{j=1}^n x_j \right)
\delta^{(2)} \left(\sum_{j=1}^n
\mathbf{k}_{\perp i}
 \right) |\psi_{n/h}(x_i, \mathbf{k}_{\perp i})|^2=1~.
\end{align}
Since  in the present analysis  we shall not investigate polarization effects, i.e.
we will evaluate unpolarized distributions, we omit here the helicity dependence. Moreover, the AdS mode functions are obtained in terms
of  the coordinates of a  $5$-dimensional space, therefore
 it is useful to rewrite the above condition in terms of
$ \tilde
\psi_{n/h}(x_i, \mathbf{b}_{\perp i})$, i.e.
 the Fourier
Transform (FT)
 of $\psi_{n/h}(x_i, \mathbf{k}_{\perp i})$, which results in

\begin{align}
\label{Eq:FT}
 \psi_{n/h}(x_j, \mathbf{k}_{\perp j}) = (4 \pi)^{(n-1)/2} \prod_{i=1}^{n-1}
d^2 \mathbf{b}_{\perp i} exp{\left(\displaystyle i
 \sum_{u=1}^{n-1} \mathbf{b}_{\perp i}
\cdot \mathbf{k}_{\perp i}   \right) } \tilde
\psi_{n/h}(x_i, \mathbf{b}_{\perp i})~,
\end{align}
where $\mathbf{b}_{\perp i}$ is the conjugate variable to $\mathbf{k}_{\perp i}$
and represents the frame independent intrinsic coordinate.
The normalization for the LF wf in coordinate space reads,

\begin{align}
 \sum_{n=2}^{\infty} \prod_{i=1}^n
 \int dx_i d^2\mathbf{b}_{\perp i} |
\tilde
\psi_{n/h}(x_i, \mathbf{b}_{\perp i})|^2=1~.
\end{align}

\subsection{The pion LF wave function}

The procedure to relate the mode functions and the LF wf can  obtained by comparing  the Drell-Yan-West
form factor \cite{Drell:1969km} with the AdS one, see
Ref. \cite{Brodsky:2007hb}. The AdS ff can be 
written from the overlap of the normalizable modes of the outgoing and incoming hadrons, $\Phi_{out}(z)$ and $\Phi_{in}(z)$ with the mode dual to the external electromagnetic source $J(Q^2,z)$,

\begin{align}
F(Q^2) = \int \frac{dz}{z^3}~e^{-\varphi_0(z)-\varphi_n(z)} \Phi_{out}(z) J(Q^2,z)\Phi_{in}(z),
\label{ff}
\end{align}
where $J(Q^2,z)$ is the bulk-to-boundary propagator. As it will be also discussed in the form factor section of this study, 
the Green's function of a vector field equation,  which has $M_5^2 =0$, is the same for the GSW model as for the SW one \cite{Brodsky:2007hb} as shown
in Refs. \cite{FolcoCapossoli:2019imm,Rinaldi:2021dxh}.
 Moreover, for large momenta the SW result coincides with that 
obtained within the HW model, and therefore we can apply the relation obtained in Ref. \cite{Brodsky:2007hb}. 
This last statement is because for large momenta the two form factors coincide since the dilaton dependence dies out. 

To establish the connection between the two form factors, the mode-function, solution of the Schr\"odinger equation, is normalized
as follows:

\begin{align}
 \int dz~ \chi(z)^2 = 1~.
\end{align}
Moreover from 
Eq. (\ref{ff}) one gets the normalization of the mode function:

\begin{align}
 \int dz~ \dfrac{e^{-\varphi_0(z)-\varphi_n(z) } }{z^3} \Phi(z)^2=1~.
\end{align}
The ``density''distribution for the pion can be obtained from the mode functions and is  related to 
the FT of the pion form factor~\cite{Brodsky:2007hb},

\begin{align}
 \tilde \rho(x,z) = \dfrac{e^{-\varphi_0(z) -\varphi_n(z)} }{2 \pi} \dfrac{x}{1-x} 
\dfrac{ |\Phi(z)|^2 }{z^4} =  \dfrac{x}{1-x} 
\dfrac{ |\chi(z)|^2 }{2 \pi z}  =
 \dfrac{|\tilde \psi_{LF}(x, \vec b_\perp) |^2}{(1-x)^2}~,
\end{align}
where in the intermediate step use has been made of Eq. 
(\ref{Eq:TR2}) for $z$ dependence. Furthermore, $x$ represents the longitudinal momentum fraction carried by a parton in the pion and
$\mathbf{ b}_\perp$ represents the transverse distance between the quark and the anti-quark.
Finally, the  relation between the pion LF wf and the mode function becomes:

\begin{align}
 \tilde \psi_{LF}(x, \mathbf{b}_\perp) \equiv \tilde \psi_{2/\pi}(x, \mathbf{b}_\perp)
=
\sqrt{\dfrac{x(1-x) }{2 \pi z}  } \chi(z)~,
\label{AdsLF}
\end{align}

where $z =\sqrt{x(1-x)} |\mathbf{b}_\perp|$ \cite{Brodsky:2007hb}. This equation allows us to obtain the  LF wf from the solution of our mode equation, 
namely the Schr\"odinger equation associated to potential  $V_\pi(z)$ which appears in Eq.(\ref{pionpotential}).

\section{The longitudinal dynamics}
\label{longitudinal}
A promising procedure to incorporate the chiral symmetry breaking in those models that have zero mass ground state pions, like that developed in Section \ref{pion},  is to include  longitudinal dynamics ~\cite{tHooft:1974pnl,Li:2015zda,Burkardt:1997de,Li:2021jqb,deTeramond:2021yyi}.
To this aim we incorporate in our scheme the procedure developed in Ref. \cite{Li:2021jqb} and also applied to the SW model of Ref. \cite{Brodsky:2007hb}. In the LF AdS/QCD framework,
the QCD hamiltonian for the pion is effectively described by 
equations such as Eqs.(\ref{psmeq}) and (\ref{pionpotential}), which depend only on the transverse coordinates $\mathbf{ z} = \sqrt{x (1-x)}
\mathbf{b}_\perp$,

\begin{align}
  -\frac{d^2}{dz^2} + V_\perp(z)   \phi(z) =M^2_\perp \phi(z)~.
\end{align}
Now we  specify the perpendicular index, since the solution to the above equation represents  the underlying transverse  dynamics of the meson. 
 However, the full description of pion structure and its spectroscopy
requires an effective way to include the chiral symmetry
breaking mechanism. To this aim we include 
longitudinal degrees of freedom in the GSW model, by following the 
line of thought of Ref. \cite{Li:2021jqb}.  Such a goal can 
be reached by assuming that the full potential is a combination of two potentials, a transverse and a longitudinal,
$V_{eff} = V_\perp+V_{||}$, and therefore the spectrum is given by two contributions to the mass
$M_\pi^2 = M^2_\perp +M_{||}^2$, determined by the equation,

\begin{align}
 \left[-\dfrac{d^2}{dz^2} + \dfrac{m_q^2}{x} + \dfrac{m_{\bar q}^2}{1-x}+
V_{eff}     
\right] \bar \Phi(z,x) = M^2 \bar \Phi(x,z)~.
\label{Eq:full}
\end{align}
The transverse quantities can be evaluated from holographic models, 
and  the longitudinal wf, and its corresponding spectrum component, can be obtained from the Schr\"odinger equation,

\begin{align}
\left[  \dfrac{m_q^2}{x} + \dfrac{m_{\bar q}^2}{1-x}+V_{||}(x) \right] 
\Xi(x)&=M^2_{||}~ \Xi(x)~,
\label{Eq:long}
\end{align}
where he $\Xi(x)$ is the longitudinal wf. In this approach 
the overall wf is given by

\begin{align}
\bar \Phi(x, \mathbf{b}_\perp) = 
\tilde\psi_{LF}(x,\mathbf{b}_\perp)~ \Xi(x)~.
\label{LFwf}
\end{align}
From now on, 
$\tilde \psi_{2/h}(x, \mathbf{b}_\perp)$ in Eq. (\ref{Eq:FT}) will be used for the { transverse} part { of the above expression in order} to 
evaluate the observables.
In order to obtain a suitable $\Xi(x)$
several longitudinal potentials have been proposed~
\cite{Li:2021jqb,Li:2015zda,Li:2022izo}. 
In this analysis we use

\begin{align}
 V_{||}(x)=-\sigma^2 \partial_x \big[x(1-x)\partial_x \big]~,
 \label{longpot}
\end{align}
due to its remarkable 
predicting power.
 Here $\sigma$ characterizes the strength of 
confinement~\cite{Li:2021jqb}. 
From this potential, a solution to  Eq. (\ref{Eq:long})
and consequently to  
Eq. (\ref{Eq:full}), can be found. The global mass spectrum is given by,
\begin{align}
 M^2 = M^2_{\perp \pi}+ \sigma (m_q+m_{\bar q} )(2l+1)+(m_q+m_{\bar q} )^2+ 
\sigma^2l(l+1)~,    
\label{mpilong}
\end{align}
and the longitudinal wf by,
\begin{align}
\Xi(x)=N_2 x^{\beta/2} (1-x)^{\alpha/2}P_l^{(\alpha,\beta)}(2x-1)~,
\label{Eq:chi}
\end{align}
where $M^2_{\perp \pi}=M^2_\pi$ in Eq. (\ref{mpi}), 
$\alpha = 2 m_q/\sigma$ and $\beta = 2 m_{\bar q}/\sigma$~
\cite{Li:2021jqb}.

In the present framework, leading order in the
$N_C$ expansion, the quarks are essentially constituent quarks
 and therefore their masses does not have to correspond to the current quark masses 
\cite{Brodsky:2007hb,Li:2021jqb}.
For the moment being,
we assume all the light quarks to have the same mass and therefore 
 $m_q = m_{\bar q}$ for any pion. Moreover
$l$ is a quantum number related to the longitudinal direction
 \cite{Li:2021jqb} and $P_l^{(\alpha,\beta)}(x)$ is a Jacobi 
polynomial. 

Fitting the pion spectrum  one finds that ground state is 
obtained for 
 $n=l=0$, the first 
excitation results from the superposition of the
 $(n=0,l=2)$ and $n=1,l=0$ states, while the second excitation is a superposition of  the  $(n=0,l=4)$, $(n=2,l=0)$ and $(n=1,l=2)$ states~\cite{Li:2021jqb} .
From this fit a relation between  $\sigma$ and  $m_q$ arises. If one requires that
the pion ground state mass is $\sim 0.14$ GeV, then from 
Eq. (\ref{mpilong}),

\begin{align}
 \sigma = \frac{0.0196-4 m_q^2}{2 m_q}~.
\label{Eq:sigma}
\end{align}

An important achievement of this approach is  that the relative parton distribution function (PDF) has the correct power
behavior $x^\alpha (1-x)^\beta$, see for example
the recent Ref.~\cite{Barry:2022itu}. We will test this behavior in the next section by 
using  the parameters that reproduce other pion observables. 

Although for the moment being the longitudinal dynamics is implemented to build
up a realistic model for the structure of the pion, this 
approach leads to remarkable description of 
meson spectroscopy \cite{Li:2021jqb}. In the present analysis,
we mainly focus on the pion spectrum and structure.

\section{Pion observables}
\label{observables}
Having described the model and found its mode function, and having established a connection 
between the AdS/QCD formalism and the Light Cone one,
we proceed now to calculate various observables and to compare with the corresponding data.
We recall that the only remaining free parameters are $\gamma_\pi$ and $m_q$.
Indeed,   $\sigma$ 
is determined from $m_q$, see Eq. (\ref{Eq:sigma}).
In particular, 
  $\gamma_\pi$ is responsible for the energy scale difference 
between the pion and the $\eta$ meson and has been added to the GSW model
 to be able to describe the pion from the pseudo-scalar EoM.
 The main motivation of the present investigation is to show that with only these 
 two additional parameters the model
 is able to describe a large amount of data and phenomenological results. In 
fact, as it will be clear in the next section, the observables we 
evaluate  depend only on the two parameters in a highly non 
linear manner. In order 
to highlight the prediction power of the model, we propose two different, but close, parametrizations. In both cases, the
essential features of the observables we analyse are qualitatively well described. In particular let us denote GSWL1 the
set: $m_q = 45$ MeV and $\gamma_\pi =-0.6$ and GSWL2 the set: $m_q=52$ MeV
and $\gamma_\pi=-0.17$.  As one can see, the value of $m_q$ is similar to that 
of Ref. \cite{Li:2021jqb}.
Let us remark that since our model formally relies in the 
leading $N_C$ physics, we expect that 
 the quark masses are constituent quark masses. In the following we call the GSW model, including the longitudinal dynamics, the GSWL model.

\subsection{The pion spectrum}
As discussed in the previous sections, thanks to some modifications of
of dilaton function in the GSW model it is possible to recover  chiral 
symmetry to describe 
the pion ground state. Moreover, the longitudinal dynamics will 
be added to realize the explicit breaking of chiral symmetry. 
In Tab. \ref{piontab} 
the results of the calculations of the pion spectrum
are shown in comparison with the PDG
data~\cite{Tanabashi:2018oca,Zyla:2020zbs}. 
We also compare the same quantity with predictions of other  
approaches, e.g. in Ref. \cite{Rinaldi:2021dxh}, where the 
chiral symmetry breaking has been effectively described by 
a complicated dilaton profile function. In the present analysis, 
once the longitudinal dynamics is considered,
the pion excitations have been obtained by 
including the contribution from the states with
longitudinal quantum number  $l>0$. In fact,
as discussed in Ref. \cite{Li:2021jqb}, the $\pi(1300)$ can be 
described as a superposition of the $|n=1, l=0 \rangle$ and
$|n=0, l=2 \rangle$ excited states, and the $\pi(1800)$ as a superposition of  
the $|n=2, l=0 \rangle$, $|n=1, l=2 \rangle$ and $|n=0, l=2 \rangle$ excited states.
If intermediate unobserved states are not allowed, the GSWL1 
parametrization of the model 
leads to a very good description of the data. Also the GSWL2 
one predicts a pion spectrum very close to the experimental scenario.
Only the $\pi'$ is underestimated.
Nevertheless, as discussed in Ref. \cite{Rinaldi:2021dxh}, from the present experimental 
scenario one cannot exclude the presence of hidden states. However, in
any case
holographic models could reproduce well the spectrum by predicting the
existence of  new states.

\begin{table}[htb]
\centering
\begin{tabular}{| c | c | c | c | c | c | c|}
\hline
& $\pi^0$ & & $\pi(1300)$ & & & $\pi(1800)$    \\ \hline
PDG & $ 134.9768\pm 0.0005$ &  & $1300\pm100$ &  & & $1819\pm 10$  \\ \hline
SW \cite{Brodsky:2007hb} &   $ 0  $    & &       $ 1080 $& $  1527
$ &  & $1870$\\ \hline
Ref. \cite{Rinaldi:2021dxh} &   $ 135  $    &$943 \pm 111$&       $ 1231  \pm 133$& $  1463 
\pm151$ & $1663 \pm168$ & $1842\pm183$\\ \hline
GSWL1 &   $ 140  $  &      & $1199\pm41 $&   &   & $1800 \pm 6$\\ 
\hline
GSWL2 &   $ 140  $  &      & $1019 \pm 27 $&   &   & $1793 \pm 16 $\\ 
\hline
Ref. \cite{Li:2021jqb} & $140$  & &1520 & & &2120 \\
\hline
\end{tabular}
\caption{ \footnotesize We show the experimental result for the $\pi$ masses given by the PDG 
particle listings~\cite{Tanabashi:2018oca,Zyla:2020zbs} together with the 
results of our calculations with the  GSWL1 and GSWL2  parametrizations. The error source in the GSWL scenarios is
associated to $\alpha = 0.55 \pm 0.04$.  The 
empty cells stand for the absent intermediate states. Masses are in MeV.}
\label{piontab}
\end{table}

\subsection{The pion form factor}

As previously discussed  the pion form factor is an essential
 quantity to investigate the pion inner structure. 
Moreover, its study allows to build up a correspondence between
the mode function of the pseudo-scalar meson and its
LF wf in coordinate space \cite{Brodsky:2007hb,Brodsky:2008pf}.
The pion form factor (ff) in the GSW can be defined as a generalization of the ff in the SW obtained by assuming
minimal coupling for the photon~\cite{Brodsky:2007hb,Polchinski:2002jw,Hong:2004sa} and it leads to

\begin{align}
      i g_5 \int d^5x \sqrt{g} e^{-\varphi_0(z)-\varphi_n(z)}
A^l \Phi^*_{P'}(x) \overset{\leftrightarrow }{\partial}_l
\Phi_P(x)~,
\end{align}
where the $\Phi_P$ represents the mode function representing a field
propagating with momentum $P^\mu$, $P^2 = M^2$ and conventionally
 $\Phi_P(x) \sim exp[i P \cdot x] \Phi(z)$. Moreover, $A^l$ represents
the mode of an electromagnetic probe propagating in this space with 
the Minkowskian virtuality vector $q^\mu$ so that 
$Q^2 = -q^2 >0$ and is given by $A_l(z) = \varepsilon_l exp[-i Q \cdot x]J(Q^2,z)$, where $J(Q^2,z)$ is the bulk-to-boundary propagator 
\cite{Erlich2:2005qh,Huang:2007fv}. The solution for $J$ in  the SW model is~\cite{Brodsky:2007hb,Zou:2018eam}:

\begin{align}
 J(Q^2,z) = \Gamma \left(1+ \dfrac{Q^2}{4 k^2} \right) 
U \left(\dfrac{Q^2}{4 k^2},0, k^2 z^2 
\right)~,
\label{backtoboundary}
\end{align}
 where $U(a,b,c)$ is the confluent hypergeometric function. In addition,
the function can be obtained from the 5th dimensional 
lagrangian of a vector field strength \cite{Zou:2018eam},

\begin{align}
      S_V = -\dfrac{1}{2}\int d^5x \sqrt{-\bar g}e^{-\beta_V k^2z^2}
\bar g^{MP} \bar g^{NQ}F_{MN}F_{PQ} &= -\dfrac{1}{2}\int d^5x \sqrt{- g}e^{ k^2z^2(-\beta_V+\alpha/2)}
 g^{MP}  g^{NQ}F_{MN}F_{PQ}
\\
\nonumber
&=\dfrac{1}{2}\int d^5x \sqrt{- g}e^{ -k^2z^2}
 g^{MP}  g^{NQ}F_{MN}F_{PQ}~.
\end{align}
The last line is obtained since $\beta_V = 1+\alpha/2$ ~\cite{FolcoCapossoli:2019imm,Rinaldi:2021dxh}. Therefore,
the photon mode  obtained for the SW model is the same as that of 
the GSW model. We recall that such a result is due to the fact that 
the conformal mass for the vector field is $M_5^2=0$ and therefore no dependence on the GSW warped metric appears.
Thus, the pion form factor within the the GSW model has the form

\begin{align}
 F_\pi(Q^2)= \int dz~ \dfrac{e^{-\varphi_0(z)-\varphi_n(z) } }{z^3} \Phi(z)^2 J(Q^2,z)~,
\end{align}
where $J(Q^2,z)$ is the bulk-to-boundary 
propagator whose solution in the SW model is given by Eq. (\ref{backtoboundary}). 
The ff can also be described in terms of the LF wf 
\cite{Drell:1969km,West:1970av}:

\begin{align}
\label{Eq:ff}
      F_\pi(Q^2) = \int  dx \dfrac{d\mathbf{k}_{\perp 1}}{16 \pi^3}~
\psi_{2/\pi}(x, \mathbf{k}_{\perp 1})\psi^*_{2/\pi}(x, \mathbf{k}_{\perp 1}+(1-x)\mathbf{q}_\perp)~,
\end{align}
where the photon 4 momentum $q^\mu=(q^+,q^-,\mathbf{q}_\perp)$
is chosen to have $q^+=0$ and $q^2=-Q^2=-\mathbf{q}_\perp^2$. 

\begin{figure*}
\includegraphics[scale=0.47]{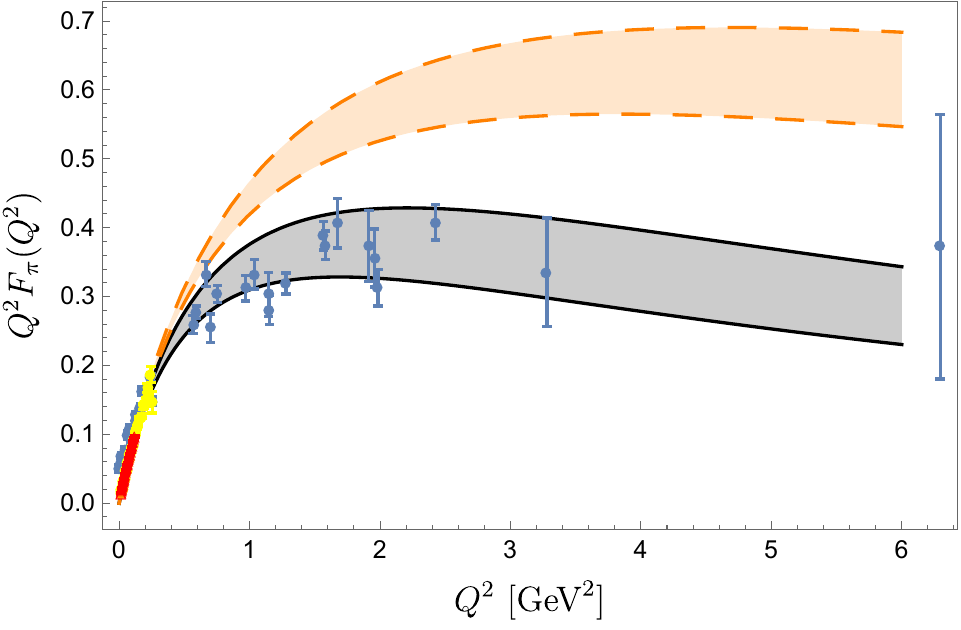}
\hskip 0.5cm 
\includegraphics[scale=0.59]{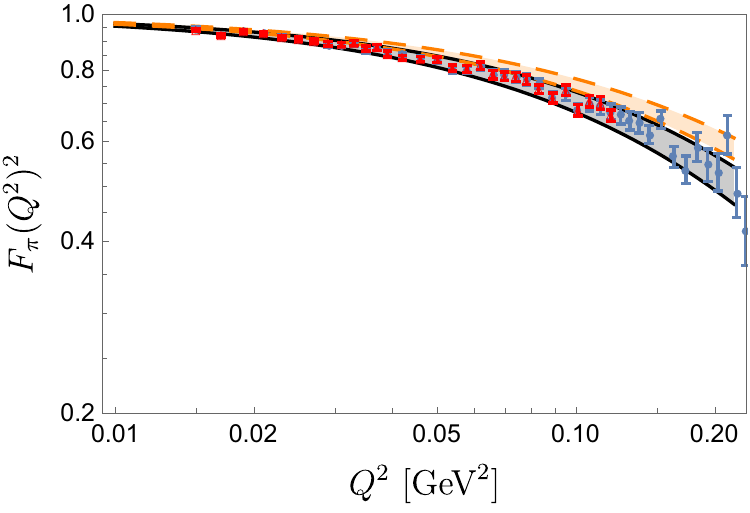}
\caption{\footnotesize  {The pion ff Eq. (\ref{Eq:ff}). Full lines for the GSWL2 model and dashed 
lines for the GSWL1 model. Data from Refs. 
\cite{NA7:1986vav,Ackermann:1977rp,Bebek:1977pe,JeffersonLabFpi:2007vir,Brauel:1979zk,JeffersonLabFpi-2:2006ysh,Amendolia:1984nz}.
 Left panel
 for $Q^2 F_\pi(Q^2)$. Right panel for $|F_\pi(Q^2)|^2$.
Bands
stand for the error in $\alpha$.  
}}
\label{fig:ff}
\end{figure*}

In Fig. \ref{fig:ff} the pion ff evaluated within the GSWL1 and GSWL2 
parametrizations is shown. In the left panel the quantity
$Q^2 F_\pi(Q^2)$ is displayed. As one can see the GSWL2 model is able 
to  reproduce quite well the ff in a wide range of $Q^2$. On the contrary,
the GSWL1 parametrization matches the data  in the low $Q^2$ region.
In order to better appreciate the comparison between the two, in the 
right panel of the figure, we show the quantity $|F_\pi(Q^2)|^2$ 
for $Q^2\leq 0.2$ GeV$^2$. In this case, both models are able to 
describe the data. We remark that the error band is due to the 
theoretical uncertainty in the parameter $\alpha = 0.55 \pm 0.04$ describing
the scalar meson spectrum within the GSW model 
\cite{Rinaldi:2020ssz}. Let us 
stress that the main  difference between the GSWL1 and GSWL2 
characterizations are due to the distinct values of $\gamma_\pi$ which
determine the energy scale of the pion spectrum. Fitting simultaneously the pion spectrum including the excitations
and the ff is difficult for most models. However, the GSW model is able to do so with the two parametrizations in the low $Q^2$ square
region.

 In the future the GSW model could
be improved  by considering  other deformations of the metric 
and/or modifications of the bulk-to-boundary propagator in order to achieve better fits. For example, in Ref. 
\cite{MartinContreras:2021yfz} the authors proposed 
a dilaton profile function for the bulk-to-boundary propagator that is proportional 
to the photon virtuality.

\subsection{The pion mean radius}
A crucial quantity encoding 
the non perturtbative structure of the pion is the charge radius.
This quantity  can be extracted from the ff or directly
from the LF wave function.
In general one defines the mean square radius by,

\begin{align}
 \langle r^2 \rangle_\pi = -6 \dfrac{d F_\pi(Q^2)}{d Q^2} \Big|_{Q^2=0}~.
\end{align}

As one can see in Tab. \ref{tabr}, both  GSWL models
are able to  reproduce, within the uncertainty on $\alpha$, the data and the most recent extractions 
of the mean pion radius
 \cite{ParticleDataGroup:2014cgo,Cui:2021aee}.
The two parametrization are almost equivalent for this 
observable.

\begin{table}
        \centering
        \label{1}
        \begin{tabular}{|c|c|c|c|c|c|c|c|}
            \hline
             & Ref.  & Ref.  & Ref.  &GSWL1&GSWL2&
Experiment& Work of Ref. \\
& \cite{Brodsky:2007hb} &\cite{Ahmady:2018muv}&  \cite{deTeramond:2018ecg}&  &   & \cite{ParticleDataGroup:2014cgo}&\cite{Cui:2021aee}\\
            \hline  
            $\sqrt{\langle r^2 \rangle}$ [fm] & 0.524&0.673-0.684 & 
0.644 &  $0.67 \pm 0.03$  &$0.70 \pm 0.05 $&0.67 $\pm 0.01$&$0.640 \pm 
0.007$ \\
            \hline          
        \end{tabular}
        \caption{ \footnotesize Values of the  pion mean radius 
obtained within different holographic models. Experimental data are from Ref. 
\cite{ParticleDataGroup:2014cgo}. In the last column we report  the recent analysis of Ref. \cite{Cui:2021aee}. }
\label{tabr}
\end{table}

\subsection{The pion effective form factor}
The pion effective form factor (eff) has been considered lately as a test for models \cite{Rinaldi:2020ybv}. Such a quantity has been 
introduced for the first time    in the context of double parton scattering
(DPS) processes in proton-proton collisions and double parton distribution 
functions (dPDFs) \cite{Rinaldi:2015cya}. 
An overview about the study of DPS processes can be found in the seminal book  Ref. \cite{Bartalini:2018qje}. In DPS two partons
of an hadron interact with two partons of the other colliding hadron.
For a long time, in order to estimate the DPS cross-section without
any phenomenological information on dPDFs,  a factorization ansatz
of  these 
quantities have been assumed for these quantity. Thanks 
to this strategy the DPS cross-section could be estimated from the 
the   product of the two single parton scattering (SPS) cross-sections scaled by an 
almost theoretical  unknown quantity called effective cross-section
\cite{Calucci:1999yz}:

\begin{align}
\label{Eq:sigmadps}
      \sigma^{A+B}_{DPS} \propto \dfrac{\sigma^A_{SPS}\sigma^B_{SPS}}{\sigma_{eff}} ~,
\end{align}
where $\sigma^{A+B}_{DPS} $ is the DPS cross-section for the production of two final states $A$ and $B$ respectively, and $\sigma^{A(B)}_{SPS} $ is the SPS cross-section for 
the production of the final state $A(B)$. 
Within this scheme, the effective cross-section, $\sigma_{eff}$, can 
be evaluated as follows:

\begin{align}
      \sigma_{eff} = \dfrac{1}{\displaystyle \int
 \dfrac{d\mathbf{k}_\perp }{(2 \pi)^2} F_{2\pi}(k_\perp^2)^2  }~,
\end{align}
where $F_{2 \pi}(k_\perp)$ is the eff and $\mathbf{k}_\perp$
is the conjugate variable to the transverse distance between 
two partons in the hadron. The eff
can be obtained as the first moment of the dPDF 
\cite{Rinaldi:2018slz}, analogously to 
the form factor that can be obtained from  the moments of the  generalized parton distribution function \cite{Diehl:2003ny}.
In terms of the LF wf of the pion, the eff is defined as follows,

\begin{align}
\label{ffc}
 F_{2 \pi}(k_\perp) &=  \int_0^1 dx~ \int {d^2 {\bf k}_{\perp 1} \over 16 
\pi^3 } ~
\psi_{2/h}^* 
(x,{\bf 
k}_{ \perp 1}  ) \psi_{2/h} (x,{\bf 
k}_{ \perp 1} +{\bf k_\perp} )
\\
\nonumber
&= \int_0^1 dx~ \int {d^2 {\bf b_\perp}  }~ 
|\tilde \psi_{LF}(x, \mathbf{b}_\perp  )|^2 e^{i \mathbf{k}_\perp \cdot 
\mathbf{b}_{\perp}  }
~.
\end{align}

In Ref. \cite{Rinaldi:2018slz}, the connection between the eff and the geometrical
 properties of the parent hadron have been established. In fact, the eff of the pion can be
related to the mean transverse distance between two partons:

\begin{align}
 \label{2dist}
  \langle b^2_\perp \rangle 
\simeq 
-4 { d 
F_{2 \pi}(k_\perp) \over 
dk_\perp^2} \Bigg|_{k_\perp=0}~,
\end{align}
where $k_\perp$ is the conjugate variable to $b_\perp$ and represents
the momentum unbalance between the first 
and the second parton in the initial and final states in  DPS 
processes.
One of the first calculations of the pion dPDF 
is that of Ref. \cite{Rinaldi:2018zng} where the SW model of Ref. \cite{Brodsky:2007hb} has been used. Thereafter a first evaluation of two-current correlations in the pion within lattice QCD
 \cite{Bali:2018nde} was carried out. Other studies  of the pion 
eff and $\sigma_{eff}$ have been discussed in Refs.
\cite{Rinaldi:2020ybv,Courtoy:2020tkd,Courtoy:2019cxq,Broniowski:2019rmu}. 

Let us mention in 
particular the study of Ref.
 \cite{Rinaldi:2020ybv}, where  the pion effs
evaluated with different holographic models have been compared 
with lattice QCD predictions in the allowed regions of $k_\perp^2$.
Indeed, lattice calculations were performed in the pion rest frame \cite{Bali:2018nde} and the eff 
has been parametrized as follows:

\begin{align}
 \label{latticeff2}
 F_{2 \pi}(k_\perp)=
 \dfrac{1}{\left[ 1+\langle 
b^2 \rangle \dfrac{k_\perp^2}{6n}  \right]^n}~.
\end{align}
A good fit to lattice data was obtained for
 $n=1.173$ and the extraction of the mean distance between two 
partons in the pion is
{ $\sqrt{\langle b^2 
\rangle} =  \sqrt{3/2 \langle b_\perp^2 \rangle}=
 1.046 \pm 0.049$ fm \cite{Bali:2018nde}}. 
 This important result  has been used to test  holographic models of the pion \cite{Rinaldi:2020ybv}. 
 As stressed in Ref. \cite{Rinaldi:2020ybv}, lattice  
outcomes
have been
obtained in the pion rest frame.
 However, the  comparison with holographic LF calculations is 
allowed in the Infinite Momentum Frame (IMF)  which can be emulated
by the kinematic condition: $k_\perp^2 <<  m_\pi^2 \sim 0.3^2$ GeV$^2$.

\begin{figure*}[htb]
\begin{center}
\includegraphics[scale=0.6]{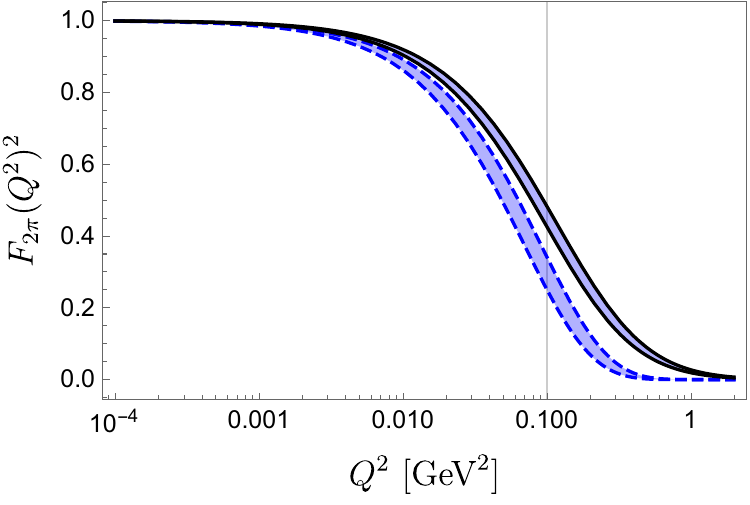}
\hskip 0.5cm 
\includegraphics[scale=0.6]{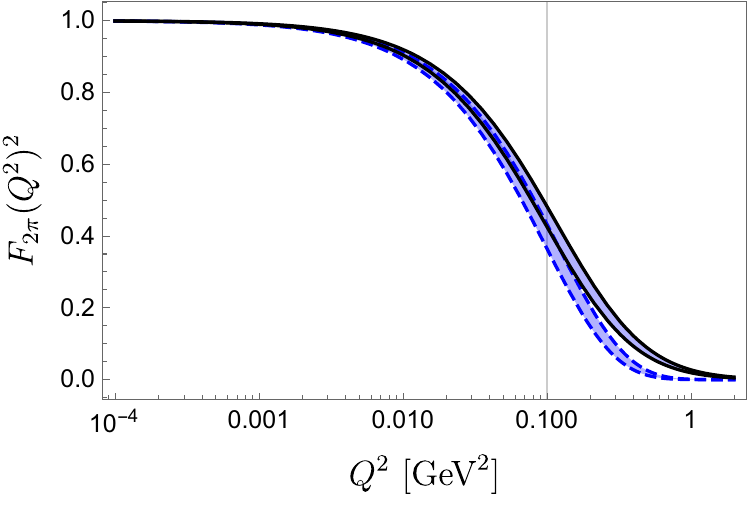}
\caption{\footnotesize  {The square of the eff Eq. (\ref{ffc}). Full lines represent lattice data \cite{Bali:2018nde} and the band stands for the error in the mean distance  entering the lattice pasteurization, see Eq. (\ref{latticeff2}). Left panel the same observable for the GSWL2. Right panel the same for the GSWL1.  }}
\label{fig:eff2}
\end{center}
\end{figure*}

\begin{figure*}[htb]
\begin{center}
\includegraphics[scale=0.88]{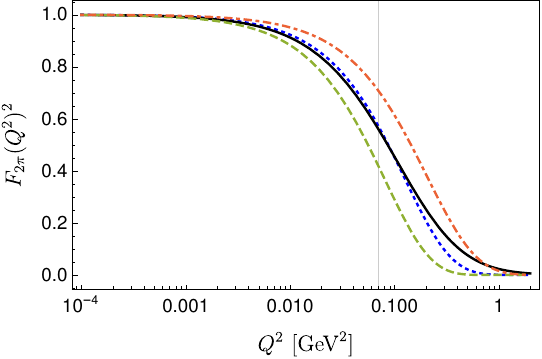}
\caption{\footnotesize  {Same of Fig. \ref{fig:eff2} from 
Ref. \cite{Rinaldi:2020ybv}. Full line the lattice calculation. Dot-dashed line calculation for the model of Ref. \cite{deTeramond:2018ecg}. Dotted line for the model of Ref. \cite{Brodsky:2007hb} and dashed line for that of Ref. \cite{Ahmady:2018muv}.}}
\label{fig:eff3}
\end{center}
\end{figure*}

In Ref. \cite{Rinaldi:2020ybv} it has been thoroughly discussed the
 difficulties in describing the ff and the eff with the same parameters in a given model. Such 
a result led to the conclusion that more sophisticated models
are necessary to reproduce the ff and the eff. 
In the present analysis,  the GSWL1 and GSWL2 parametrization were 
used to evaluate
Eqs. (\ref{ffc}, \ref{2dist}). 
In Fig. \ref{fig:eff2} 
the square of the pion eff is displayed showing the comparison between lattice data and our model calculations. Let us stress that 
since the DPS cross-section, even in the simplified description of 
Eq. (\ref{Eq:sigmadps}) depends on the square of the eff, in
 Fig. \ref{fig:eff2} we report such a quantity, instead of the eff, in order
to highlight the relevant differences that could affect the evaluation of experimental observables, such as $\sigma_{eff}$. As one can see, in 
this case the GSWL1 model is able to reproduce the lattice data with 
impressive accuracy. On the other hand the GSWL2, although it provides a reasonable agreement, it underestimates slightly this 
quantity.  We show in Fig. \ref{fig:eff3}, extracted from Ref. \cite{Rinaldi:2020ybv}, the results of other model calculations. It can be noted, that despite the limited 
success of the SW model in describing the pion ff, such a model is 
capable of describing the eff lattice data. Such feature is shared 
with the GSWL1 parametrization, where the ff is not overall well 
reproduced but the eff is close to lattice data. On the contrary
the GSWL2 model, which  describes well the experimental data of the ff, underestimates the eff lattice data. 
 Finally,  in Tab. \ref{tabd} we compare
the mean distance between two partons in the pion evaluated in 
different models with the lattice predictions.
In this case, the pion GSWL1 parametrization, without any dedicated parameter, is the only model that reproduces the lattice data within
error. Also the GSWL2  predicts a value for this quantity close
to the data. 

The calculation of the eff confirms again the difficulty 
in reproducing the mean pion charge radius and the mean distance between two partons in 
the pion with the same parameters, . However, at variance with the SW model which leads to 
a good value for this distance but fails in evaluating 
$\sqrt{\langle r^2 \rangle }$, the GSWL1 is able to reproduce 
simultaneously $\sqrt{\langle r^2 \rangle }$ and $\sqrt{\langle b^2 \rangle }$. Coherently with the eff analysis, the 
GSWL2 is not able to reproduce the mean distance with
an error very close to that of the SW calculation.
However, the GSWL2 model calculation of the pion charge radius matches  
the data.

In closing this section the analysis of the data so far of all the holographic models allows us to conclude that the GSWL1 and GSWL2 parametrizations are capable of describing reasonably well the one-body and two-body ffs simultaneously. From this  point of view the graviton soft-wall model including
the longitudinal dynamics is very promising in describing the DPS pion physics.

\begin{table}
        \centering
        \label{1}
        \begin{tabular}{|c|c|c|c|c|c|c|}
            \hline
             & Model of Ref.  & Model of Ref.&Model of Ref.&Lattice&GSWL1& GSWL2\\
& \cite{Brodsky:2007hb} &\cite{Ahmady:2018muv}&  \cite{deTeramond:2018ecg}& \cite{Bali:2018nde}& &\\
            \hline  
            { $\sqrt{\langle b^2 \rangle}$ [fm]} & 0.968&1.207 & 
0.767 &1.046 $\pm 0.049$ &$1.13 \pm 0.06$& $1.24 \pm 0.08$ \\
            \hline          
        \end{tabular}
        \caption{\footnotesize Values of the mean partonic distance 
in the pion, Eq. 
(\ref{2dist}), 
obtained in lattice QCD \cite{Bali:2018nde} and in models based on the AdS/QCD 
approach \cite{Brodsky:2007hb,Ahmady:2018muv,deTeramond:2018ecg}. }
\label{tabd}
\end{table}

\subsection{The pion decay constant}
In this section results for the calculation of 
the pion decay constant are presented and discussed. 
This quantity can be defined from the parametrization of the Lorentz
structure of the following amplitude:

\begin{align}
 \langle 0|\bar \psi \gamma^+ \dfrac{1}{2}(1-\gamma_5) \psi
|\pi \rangle = i \dfrac{P^+ f_\pi}{\sqrt{2}}~.
\end{align}
Thanks to the LF wf representation of the pion state, leading to the Fock 
expansion previously described,  one can relate the pion 
decay constant to its LF wf \cite{Lepage:1980fj,Brodsky:2007hb}:

\begin{align}
 f_\pi = 2 \sqrt{N_C} \int_0^1 dx~\int \dfrac{d \mathbf{k}_{\perp 1}}{16 \pi^3}  
\psi_{2/h}(x, \mathbf{ k}_{\perp 1})~.
\end{align}
In order to avoid confusion with other definitions, used in  e.g. 
Ref. \cite{Mondal:2021czk}, let us specify that within the present formalism we 
consider the experimental data \cite{Tanabashi:2018oca}
 $(130 \pm 5)/\sqrt{2}=91.92 \pm 3.54$ MeV.
 Comparisons with  data  and other model calculations
 \cite{Li:2021jqb,Brodsky:2007hb,Ahmady:2018muv}
 are displayed in Tab. \ref{tab2}. We stress again that 
the quoted value in that references might be rescaled by a factor $1/\sqrt{2}$
when needed.
As one can see, holographic approaches are qualitatively 
able to describe the decay constant. In order to highlight the
discrepancy with the data and model predictions, the quantity 
$\Delta_\pi = |f_\pi^{holo}-f_\pi^{exp}|$ is shown in 
 Tab.  \ref{tab2}. As one can see the GSWL2 parametrization and the model of
 Ref. \cite{Ahmady:2018muv}
is quite close to the data. In this case, a bigger 
discrepancy is found for the GSWL1 case, although the result is not too distant from the data.

\begin{table}[h]
\begin{center}
\begin{tabular}{c|ccccccc}
\hline
\hline
     \rule{0mm}{0.5cm} 
&Data \cite{Tanabashi:2018oca}&GSWL1 &GSWL2 &Work of Ref. \cite{Li:2021jqb}& SW 
\cite{Brodsky:2007hb}
& Work of Ref. \cite{Ahmady:2018muv}
\\    
\hline
\rule{0mm}{0.5cm} 
$f_\pi$ [MeV] &$91.92 \pm 3.54$& $126 \pm 6$ &$104 \pm 7$ & 
129 & 81.2 &95.5-97.6
\\    
\hline
\rule{0mm}{0.5cm} 
$\Delta_\pi$ [MeV] & & $34\pm 7$ & $12\pm 8$
 &$37.1\pm 3.5 $ & $11\pm 4$&$3.1\pm 3.5-5.7\pm 3.5$
\\
\\
\hline
\hline 
\end{tabular}
\caption{Comparison of the pion decay constant between holographic models \cite{Brodsky:2007hb,Ahmady:2018muv} 
and data \cite{Tanabashi:2018oca}. The difference in modulus between the experimental value and the model predictions is display in the last line.}
\label{tab2}
\end{center}
\end{table}

\subsection{The pion distribution amplitude}
In this section we provide the calculation of the pion distribution amplitude
(DA)
with the Light-Front formalism for the GSWL models.  There are many for DA and we choose the one of Ref. ~\cite{Brodsky:2007hb,Lepage:1980fj}

\begin{align}
\label{Eq:DA}
  \phi(x;Q) =  \int_0^{Q^2} \dfrac{d^2 \mathbf{k}_{\perp 1}  }{16 \pi^3} 
\psi_{2/\pi}(x,\mathbf{k}_{\perp 1} )~.
\end{align}
We stress that, in analogy with the SW results, also in the  GSWL models the soft $Q^2$ dependence, 
due to the wf dependence on the transverse momentum, can be safely 
neglected for $Q^2>1$ GeV$^2$. 
Thanks to this choice, the present evaluation can be compared to 
the  asymptotic expression for the DA~\cite{Lepage:1980fj,Brodsky:1984xk,Radyushkin:1977gp}:

\begin{align}
  \phi^{asy}_{pQCD}=\sqrt{3} f_\pi x(1-x)~.
\label{Eq:DA_asy}
\end{align}
 We also display the DA obtained within the SW model of Ref. \cite{Brodsky:2007hb},

\begin{align}
\label{Eq:DAasySW}
   \phi_{AdS}^{asy}(x)=4 f_\pi \sqrt{x(1-x)}/(\sqrt{3} \pi)   
\end{align}
 
\begin{figure*}[htb]
\begin{center}
\includegraphics[scale=0.4]{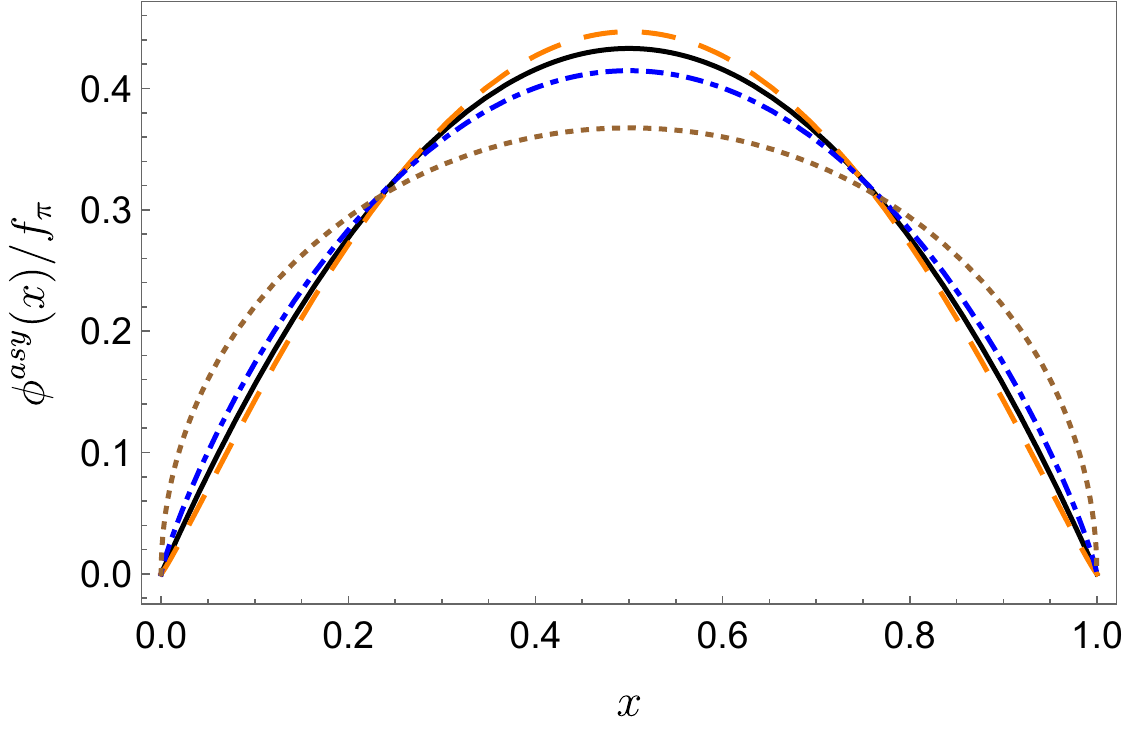}
\hskip 0.5cm 
\includegraphics[scale=0.6]{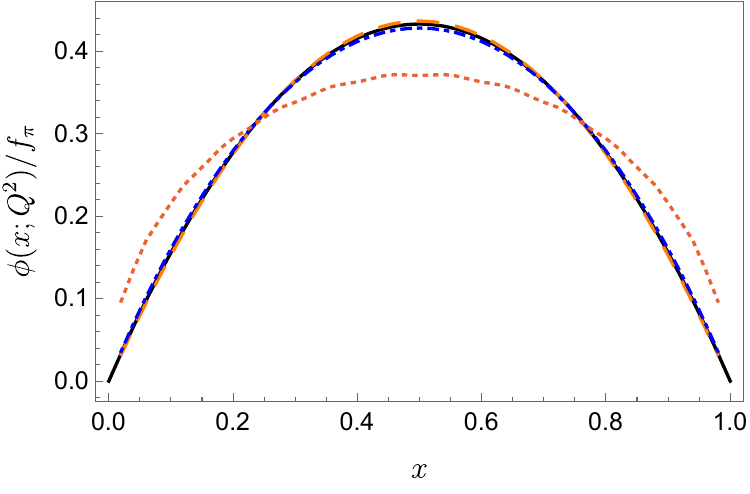}
\caption{\footnotesize  {The pion DA. Full lines represent the asymptotic pQCD
prediction  whose expression is in Eq. (\ref{Eq:DA_asy}). Dashed 
 lines for the GSWL2 parametrization. Dot-Dashed line for GSWL1 and
dotted lines for the GSW and SW \cite{Brodsky:2011yv} models, see Eq.
(\ref{Eq:DAasySW}).
Left panel: calculations of the DA in the asymptotic limit without  
evolution. Right panel the same as the left panel 
but including the ERBL evolution for $Q^2=1.5$ GeV$^2$. }}
\label{fig:DA}
\end{center}
\end{figure*}

One should notice that in the asymptotic regime, the scale dependence of the model
disappears. Therefore the DA evaluated with the SW
 model of Ref. \cite{Brodsky:2007hb} and 
discussed in Ref. \cite{Brodsky:2011yv}, is the same as that obtained with the
GSW model not including  the longitudinal dynamics.
On the other hand, the GSWL prediction for the DA, in the 
mentioned region  is

\begin{align}
\label{Eq:DAasyGSW}
  \phi^{asy}_{GSWL} \propto \big[x(1-x) \big]^{\frac{1}{2}+ \frac{m_q}{\sigma}~,  }
\end{align}
where for the
GSWL1 parametrization the exponential is 0.852 while for the GSWL2 parametrization is 1.12.
 The models which incorporate longitudinal dynamics produce an
$x$ dependence of the DA in closer agreement with the pQCD asymptotic behaviour than the SW model
of Ref. \cite{Brodsky:2007hb}. Such features can be observed in the left panel of Fig. \ref{fig:DA}.

\subsubsection{Evolution of the DA}
 As discussed in Ref. \cite{Brodsky:2011yv}, the $Q^2$
dependence of the DA has two sources: $i)$  the soft one, due to 
the integration of the wf up to the scale $Q^2$, see Eq. (\ref{Eq:DA}), and $ii)$ the hard one, due to the ERBL evolution of the DA~\cite{Lepage:1980fj,Efremov:1979qk}.
In fact, within the SW and GSW models, in general one could write 
the DA, including the soft $Q^2$ dependence as follows:

\begin{align}
\label{Eq:DAsoft}
       \phi(x;Q^2) =\phi^{asy}(x)\phi_{soft}(x,Q^2) = \phi^{asy}(x) \left[1-exp\left(-
 \dfrac{Q^2}{\kappa^2\sqrt{-2+\dfrac{1-2 \gamma}{\alpha^2}  } x(1-x) }   \right)  \right]~,
\end{align}
where, in the case of the GSW model, 
$\kappa^2 =0.37^2/\sqrt{\alpha}$ GeV$^2$.
From the above equation it is clear that from the model calculation of 
$\phi$ one gets $ \phi^{asy}$ by considering that 
$ \phi^{asy}(x) = \phi(x,Q^2 \rightarrow \infty)$.
In this case, this soft dependence is almost negligible for $Q>$1 GeV.
In order to properly take into account also the perturbative 
QCD effects, 
the distribution $\phi^{asy}(x) =  \phi^{asy}(x,\mu_0)$
 appearing in Eq. (\ref{Eq:DAsoft}) must be 
evolved by using 
the
ERBL evolution of the DA. 

We 
recall that due to ERBL evolution at a given final momentum scale the DA reads,

\begin{align}
   \phi^{asy}(x;Q) = x(1-x) \sum_{n=0,2,4...}^{\infty}a_n(Q) C^{3/2}_n (2x-1)~,
\end{align}
where here $C_n^\alpha(x)$ are Gegenbauer polynomials and 

\begin{align}
\label{Eq:Evo1}
a_n(Q) = \left[ \dfrac{\alpha_s(\mu_0^2) }{\alpha_s(Q^2)}   
\right]^{\gamma_n/\beta_0}a_n(\mu_0)~.
\end{align}
In the above expressions the following quantities appear

\begin{align}
&\beta_0 = 11- \dfrac{2}{3}n_f
\\
&  \alpha_s(Q^2)  = \dfrac{4 \pi}{\beta_0 \ln(Q^2/\Lambda_{QCD}^2)  }
\\
&\gamma_n = \dfrac{4}{3} \left[3+ \dfrac{2}{(n+1)(n+2)}-4 
\sum_{j=1}^{n+1} \dfrac{1}{j} \right]~,
\\
&a_n(\mu_0) = \dfrac{4 (2n+3)}{(n+2)(n+1)} \int_0^1 dx~
 \phi^{asy}(x,\mu_0)
C_n^{3/2}(1-2x)~,
\label{Eq:Evo2}
\end{align}
where $n_f=3$ and $\Lambda_{QCD} \sim 0.225$ GeV. 
Finally, once the the soft and hard $Q^2$ parts are separated, 
the total evolved DA
 is given by

\begin{align}
      \phi(x;Q^2) = \phi^{asy}(x;Q^2) \phi_{soft}(x;Q^2)~.
\end{align}

\begin{figure*}[htb]
\begin{center}
\includegraphics[scale=0.7]{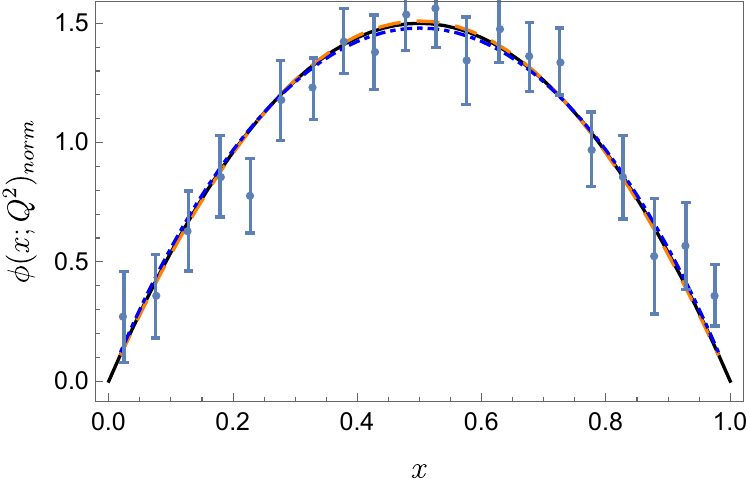} 
\caption{\footnotesize  {
Comparison between the normalized DA, see Eq.
(\ref{Eq:NormDA}) and the data from Ref. \cite{E791:2000xcx}.
The final scale is $Q=3.16$ GeV. The  full line shows the pQCD 
prediction. The dashed line corresponds to the GSWL2 parametrization
and the dot-dashed 
line is related to the GSWL1 one. }}
\label{fig:TFFR}
\end{center}
\end{figure*}

In analogy with Refs.~\cite{Mondal:2021czk,Brodsky:2011yv}, we fix the initial scale of the model 
by comparing the evolution of parton distribution functions with 
the corresponding data. In particular, 
as will be properly discussed in a next subsection, 
we consider $0.07 \leq \mu_0^2 \leq 0.11$ GeV$^2$.
In the left panel of Fig. \ref{fig:DA}, the
asymptotic
 DA
evaluated with the SW model~\cite{Brodsky:2011yv} and the 
GSWL models, is compared 
with the pQCD predictions. 
Effects of ERBL evolution of the DA are shown in the right panel of Fig. 
\ref{fig:DA}. Here we  display the DA evolved to $Q^2=1.5$ GeV$^2$ 
from  $\mu_0^2= 0.07$ GeV$^2$ for the GSWL1 and GSWL2 parametrizations.
The evolution brings the DA towards the pQCD asymptotic limit. This feature
 is due to the power of the exponential which, in the GSWL models, is 
close to the pQCD prediction already at the initial scale.  
In Fig. \ref{fig:TFFR} we compare the 
normalized DA with the data of Ref. \cite{E791:2000xcx}. As one 
can see good agreement is obtained. {  Let us specify that the 
DA appearing in Fig. \ref{fig:TFFR} can be obtained from that 
satisfying Eq. (\ref{Eq:DA}) with a simple re-scaling so that:}

\begin{align}
\label{Eq:NormDA}
      \int dx~\phi(x;Q^2)_{norm} = 1~.
\end{align}

 An important test for the DA is represented by 
the moments of the DA:

\begin{table}[t]
\begin{center}
\begin{tabular}{c|cccc}
\hline
\hline
     \rule{0mm}{0.5cm} 
&$\langle z^2 \rangle$ &$\langle z^4 \rangle$ &$\langle z^6 \rangle$ &$\langle z^{-1} \rangle$ 
\\    
\hline
\rule{0mm}{0.5cm} 
Asymptotic  &0.200 & 0.086 &0.048 & 3.000 
\\    
\hline
\rule{0mm}{0.5cm} 
Ref. \cite{Mondal:2021czk} &0.217 & 0.097 &0.055 & 3.170 
\\    
\hline
\rule{0mm}{0.5cm} 
Ref. \cite{Choi:2007yu} &$0.24 \pm 0.22$ & $0.11 \pm 0.09$ &$0.07 \pm 0.05 $ & 
\\    
\hline
\rule{0mm}{0.5cm} 
Ref. \cite{Brodsky:2011yv} 
&0.25 & 0.125 &0.078 & 3.98
\\    
\hline
\rule{0mm}{0.5cm} 
Ref. \cite{Ahmady:2018muv} &0.185 & 0.071 &  & 2.85
\\    
\hline
\rule{0mm}{0.5cm} 
Ref. \cite{Ahmady:2018muv} &0.200 & 0.085 & & 2.95  
\\
\hline
\hline
\rule{0mm}{0.5cm} 
Lattice & &  & &  
\rule{0mm}{0.5cm} 
\\
\hline
\rule{0mm}{0.5cm} 
Ref. \cite{Arthur:2010xf} &$0.28^{+0.01}_{-0.01}$ &  & &    
\\
\hline
\rule{0mm}{0.5cm} 
Ref. \cite{Braun:2015axa} &$0.2361^{+0.0041}_{-0.0039}$ &  & & 
\\
\hline
\rule{0mm}{0.5cm} 
Ref. \cite{Braun:2006dg} &$0.27 \pm 0.04$ &  & &  
\\
\hline
\rule{0mm}{0.5cm} 
Ref. \cite{Bali:2017ude} &$0.2077 \pm 0.0043$ &  & &   \\
\hline
\rule{0mm}{0.5cm} 
Ref. \cite{RQCD:2019osh} &$0.234 \pm 0.006$ &  & &   \\
\hline
\rule{0mm}{0.5cm} 
Ref. \cite{Zhang:2020gaj} &$0.244 \pm 0.030$ &  & &   
\\
\hline
\hline
\rule{0mm}{0.5cm} 
GSWL1 &$0.2033 \pm 0.0007 $ &$0.0880 \pm 0.0005$  &$0.0493\pm 0.0004$ & $3.0349 \pm 0.0087$ 
\\
\hline
\rule{0mm}{0.5cm} 
GSWL2 &$0.19770 \pm 0.0005  $ &$0.0841 \pm 0.0004$  &  $0.0465 \pm 0.0002$ & 
$2.9770 \pm 0.0055$
\\
\hline
\hline
\end{tabular}
\caption{Comparison between the moments of the DA evaluated with the 
GSWL models and other approaches. The error in the GSWL1 and GSWL2 models 
is due to the uncertainty of the initial scale $0.07 \leq \mu_0^2 \leq 0.1$.}
\label{tab3}
\end{center}
\end{table}

\begin{align}
\label{Eq:mome0}
  \langle z^p \rangle = \dfrac{ \displaystyle  \int_0^1 dx~z^p \phi(x,Q^2)}{\displaystyle  \int_0^1 dx~ \phi(x,Q^2)}~
\end{align}
where $z=2x-1$ for $p\geq 1$ and $z=x$ for $p=-1$. As one can see in Tab. 
\ref{tab3},  
the GSWL models produce results close to the pQCD 
predictions, as expected from the power behaviour of the DA,
Eq. (\ref{Eq:DAasyGSW}), as compared to that of Eq. (\ref{Eq:DA_asy}).
In all cases, the results are close to the other phenomenological models.
  Let us mention that numerically, results displayed in Tab.
\ref{tab3}, obtained within the GSWL models, are consistent with the alternative
expression of the DA moments \cite{Choi:2007yu} (see Appendix \ref{AppA}):

\begin{align}
\label{Eq:mome}
  \langle z^2\rangle &=  \dfrac{6}{a_0(Q)} \left( \dfrac{2}{35}a_2(Q)+\dfrac{1}{30}a_0(Q)  \right)
\\
\nonumber
\langle z^4\rangle &=
\dfrac{6}{a_0(Q)} \left( \dfrac{4}{231}a_4(Q)+
 \dfrac{4}{105}a_2(Q)+\dfrac{1}{70}a_0(Q)  \right)
\\
\nonumber
\langle z^6\rangle &=
\dfrac{6}{a_0(Q)} \left(\dfrac{32}{6435}a_6(Q)+
  \dfrac{20}{1001}a_4(Q)+
 \dfrac{2}{77}a_2(Q)+\dfrac{1}{1260}a_0(Q)  \right)
~.
\end{align}

\subsection{Pion-photon transition form factor}

\begin{figure*}[htb]
\begin{center}
\includegraphics[scale=0.6]{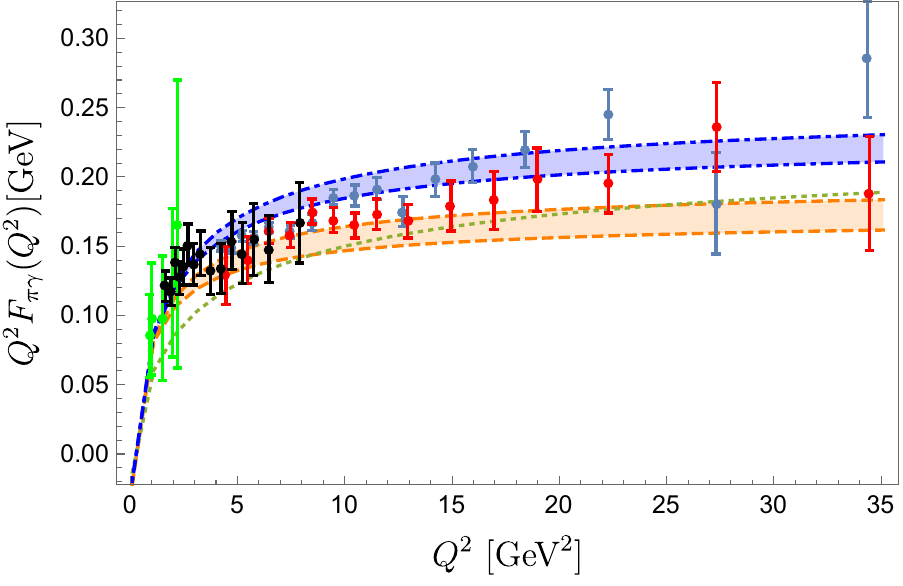}
\caption{\footnotesize  {
The pion to photon transition form factor, Eq. (\ref{Eq:TFFG}) evaluated 
within the GSWL2 (dashed) and GSWL1 (dot-dashed) parametrizations. The dotted line
stands for the SW prediction. Data are collected from Refs. 
\cite{Belle:2012wwz,BaBar:2009rrj,CLEO:1997fho,CELLO:1990klc}.
The initial scale is $\mu_0^2=0.07$ GeV$^2$.}}
\label{fig:TFFR1}
\end{center}
\end{figure*}

In this section we discuss the calculation of the pion-photon
 transition form factor. Such a quantity is relevant for processes, 
e.g., $\gamma^*(q) \gamma \rightarrow \pi^0$ and can be defined 
through the matrix element of the following electromagnetic current~\cite{Lepage:1980fj}:

\begin{align}
      \langle \gamma (P-q)|J^\mu|\pi(P) \rangle =i e^2 F_{\gamma \pi}(Q^2)
\varepsilon^{\mu \nu \rho \sigma}P_\nu \varepsilon_\rho q_\sigma~,
\end{align}
where $P$ is the pion 4-momentum, $q=-Q$ is the photon virtuality and 
$\varepsilon_\rho$
 is the polarization vector. At LO, the transition ff can be evaluated from 
the DA convoluted with a specific kernel~\cite{Cao:1996di,Musatov:1997pu,Brodsky:2011yv}:

\begin{align}
\label{Eq:TFFG}
      Q^2 F_{\gamma \pi}(Q^2) &= \dfrac{4}{\sqrt{3}}\int_0^1 dx~
T_H(x,Q^2) \int_0^{\bar Q} \dfrac{d\mathbf{k}_{\perp 1}   }{16 \pi^3}
\psi_{2/\pi}(x,\mathbf{k}_{\perp 1})
\\
\nonumber
&= \dfrac{4}{\sqrt{3}}\int_0^1 dx~
T_H(x,Q^2) \phi(x,\bar Q)~
\end{align}
where here $\bar Q = (1-x) Q$ and at LO

\begin{align}
      T^{LO}_H(x,Q^2) =  \dfrac{1}{1-x}
\end{align}
while at NLO 

\begin{align}
\label{Eq:TGNLO}
      T^{NLO}_H(x,Q^2) = T_H^{LO}(x,Q^2)\left[1- 
\dfrac{\alpha(Q^2)}{4\pi}C_f\left(9+ \dfrac{1-x}{x}\log(1-x)-
\log(1-x)^2  \right)   \right]~,
\end{align}
where further logarithms are neglected by setting the regularization 
scale to be equal to $Q$~\cite{Brodsky:2011yv,Mondal:2021czk}.
In the 
asymptotic limit one should approach the pQCD prediction:
$Q^2 F_{\gamma \pi}(Q^2) \rightarrow 2 f_\pi$~\cite{Brodsky:2011yv}.
Deviation from the 
latter condition are expected from model calculations that differ 
from the well known DA $\phi(x)\sim x(1-x)$ in the asymptotic region.
In Fig. \ref{fig:TFFR1} the transition ff, Eq. (\ref{Eq:TFFG}), is shown 
for both the GSWL reparametrizations.
To this aim the DA has been evolved  by properly taking into account the 
soft and hard $Q^2$ dependences. Such an approach has been investigated in detail
in Ref. \cite{Brodsky:2011yv}. 
The above quantity has been obtained for 
$\mu_0^2 = 0.07$ GeV$^2$. No relevant differences are found for 
$  0.07\leq \mu_0^2 \leq 0.11$ GeV$^2$.
As one can see in Fig. \ref{fig:TFFR1} the data~\cite{Belle:2012wwz,BaBar:2009rrj,CLEO:1997fho,CELLO:1990klc} 
are well reproduced. From a numerical point of view, in order to guarantee the convergence of the integral Eq. (\ref{Eq:TFFG}), we need to impose that 
for $(1-x)Q < \mu_0$ then $\bar Q = \mu_0$ ~\cite{Brodsky:2011yv,Mondal:2021czk}. 
While the SW model is not able to describe the data the GSWL  models do well.

\subsection{Virtual photon transition form-factor}
\begin{figure*}[t]
\includegraphics[scale=0.53]{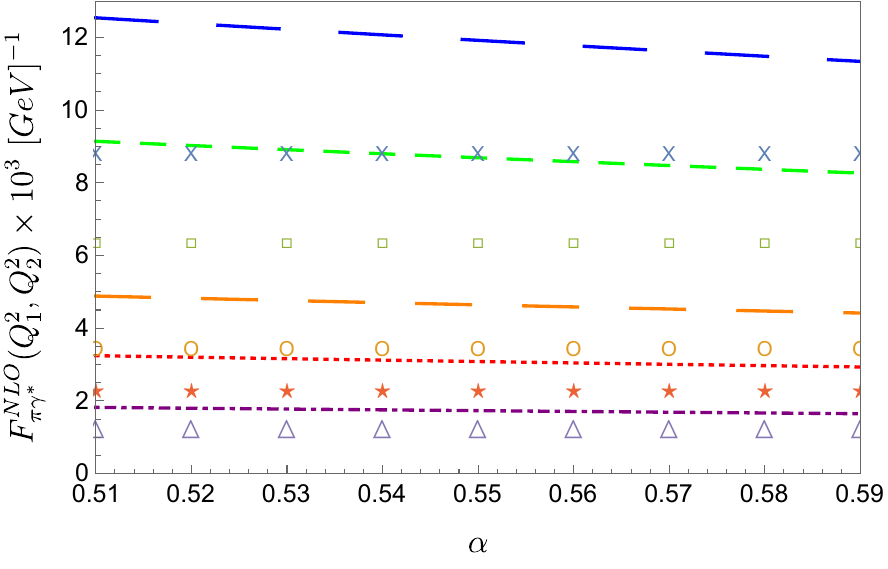} \hskip 0.2cm
\includegraphics[scale=0.53]{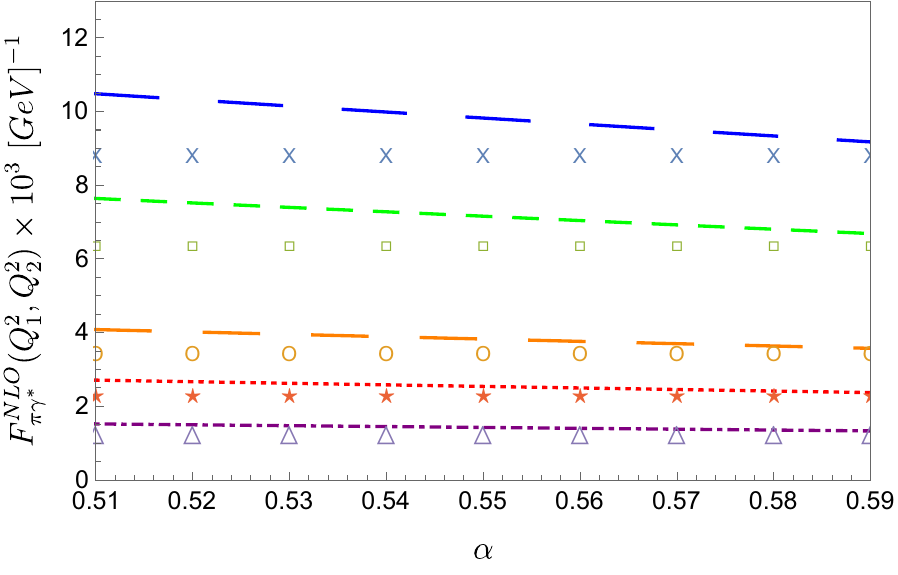}
\caption{\footnotesize  {
The pion virtual photon transition form factor evaluated at NLO
as a function of the parameter $\alpha$. Lines stand for the GSWL calculations
and markers for the pQCD predictions \cite{Mondal:2021czk}.
Long dashed blue line and blue crossed points for $(Q_1^2=Q_2^2=6.48)$ GeV$^2$.
 Middle dashed orange line and orange circle points for
$(Q_1^2=Q_2^2=16.85)$ GeV$^2$. Small dashed green line and green
 square points for $(Q_1^2=14.83;~Q_2^2=4.27)$ GeV$^2$.
Dotted red line and red star points for $(Q_1^2=38.11;~Q_2^2=14.
95)$ GeV$^2$. Dot-Dashed purple line and
purple triangle points for $(Q_1^2=Q_2^2=45.63)$ GeV$^2$. Left panel for GSWL1. Right panel for GSWL2.
 }}
\label{fig:TFFV}
\end{figure*}

\begin{figure*}[t]
\includegraphics[scale=0.55]{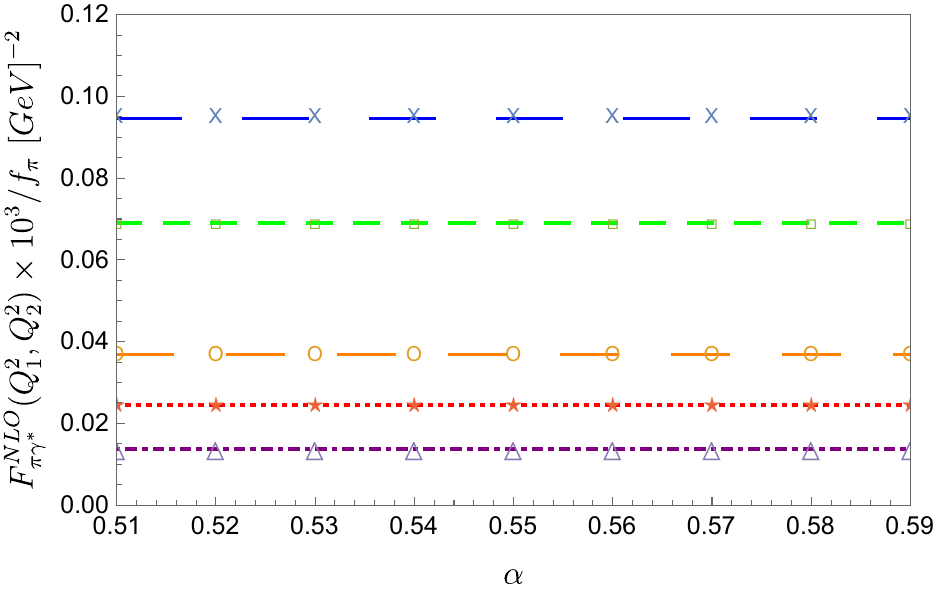} \hskip 0.2cm
\includegraphics[scale=0.55]{TFFVf_1.pdf}
\caption{\footnotesize  {
Same of fig. \ref{fig:TFFV} but for $F_{\pi \gamma^*}/f_\pi$. 
 }}
\label{fig:TFFV2}
\end{figure*}

We present also the  calculation of the transition ff for a pion
which decays in two virtual photons: $\pi^0 \rightarrow \gamma^* \gamma^*$. This quantity now depends on the virtualities of the two 
photons $Q_1,Q_2$~\cite{Lepage:1980fj,Brodsky:1981rp},

\begin{align}
       F_{\pi \gamma^*}(Q_1^2,Q^2_2) = 
 \dfrac{2}{ \sqrt{ 3}} \int_0^1 dx~ \bar T_H(x,Q_1^2,Q^2_2) \phi(x,W)~,
\label{eq:TFF2}
\end{align}
where $\bar T_H(x,Q_1^2,Q^2_2)$ is the hard-scattering amplitude for the 
present process. At LO:

\begin{align}
       \bar T_H^{LO}(x,Q_1^2,Q^2_2) = \dfrac{1}{(1-x)Q_1^2+xQ_2^2}~,
\label{Eq:kernelTFFV}
\end{align}
while at NLO
the kernel has been derived in Ref. \cite{Braaten:1982yp}
 and we show it in appendix \ref{AppB}. One should notice that the transition ff with 
a real photon is obtained for $Q_1=0$ or $Q_2=0$. In addition, at LO, for $Q_1=Q_2$, the amplitude does not depend on $x$, and therefore 
one gets $Q^2 F_{\pi \gamma^*}(Q^2,Q^2) \rightarrow 
\sqrt{2}/3 f_\pi  $.
In Fig. \ref{fig:TFFV} the results with the GSWL parametrizations are compared with
NLO pQCD calculations. As one can see the second parametrization GSWL2 provides the
expected results within the theoretical error in $\alpha$. On the other hand 
GSWL1 overestimates the pQCD result. Nevertheless, since 
in the asymptotic regions, for $Q_1=Q_2$, this ff is proportional to the decay constant, in order  to exclude from the calculations the error related to this quantity, 
  in Fig. \ref{fig:TFFV2} we plot
$F_{\pi \gamma^*}/f_\pi$. As one can see, for both the 
GSWL parametrizations, the model is able to reproduce the pQCD predictions which
are determined by both the NLO correction to the kernel Eq. (\ref{Eq:kernelTFFV}) and the ERBL evolution of the DA.
The scaling behavior $F_{\pi \gamma^*}(Q_1^2,Q_2^2) \sim 1/(Q_1^2+Q_2^2)$ is verified.

\subsection{The pion parton distribution function}
In this last section we present the calculation of the parton distribution
function  of the pion. These quantities can be defined in terms 
of the LF wf,

\begin{align}
      f(x;\mu_0^2) = \int \dfrac{d\mathbf{k}_{\perp 1} }{16 \pi^3}
|\psi_{2/\pi}(x,\mathbf{k}_\perp)|^2 = [(1-x)x]^{2 m_q/\sigma}
\dfrac{\Gamma\left[ 2+ \dfrac{4 m_q}{\sigma}   \right]}{
  \Gamma\left[ 1+ \dfrac{2 m_q}{\sigma}   \right]^2 }~.
\end{align}
Several analyses have been performed within holographic models, see
e.g. Refs.
 \cite{deTeramond:2018ecg,Ahmady:2018muv,Chang:2020kjj,Lan:2019rba}. Let us also mention predictions from QFT based models such
 those of, e.g. Refs. \cite{dePaula:2022pcb,Noguera:2015iia}.

Let us recall that, if only the LF wf determined by the modes of the pseudo-scalar field, propagating in the modified metric, are considered, the 
 PDF will be constant, as discussed for the  SW model of Ref. \cite{Brodsky:2011yv}. Therefore, in this investigation, we take advantages of 
 the longitudinal dynamics introduced in order to describe the chiral symmetry  breaking. Such feature leads to a complex structure of the 
 pion already appreciated in the study of the ff. In this case  the GSWL1 parametrization leads to the following PDF

\begin{align}
      f(x;\mu_0^2) =3.63798 [(1-x)x]^{0.704348}~
\end{align}
while the GSWL2 parametrization to the following

\begin{align}
      f(x;\mu_0^2) =8.7889  [(1-x)x]^{1.23133}~.
\end{align}
Let us remark that the GSWL2 parametrization predicts that 
$f(x \rightarrow 1) \sim (1-x)^{1.23133}$, such an exponent is close
to that found in e.g. Refs. \cite{Lan:2019rba,dePaula:2022pcb} and in
phenomenological studies such as in Ref. \cite{Courtoy:2020fex}.

\begin{figure*}[t]
\includegraphics[scale=0.6]{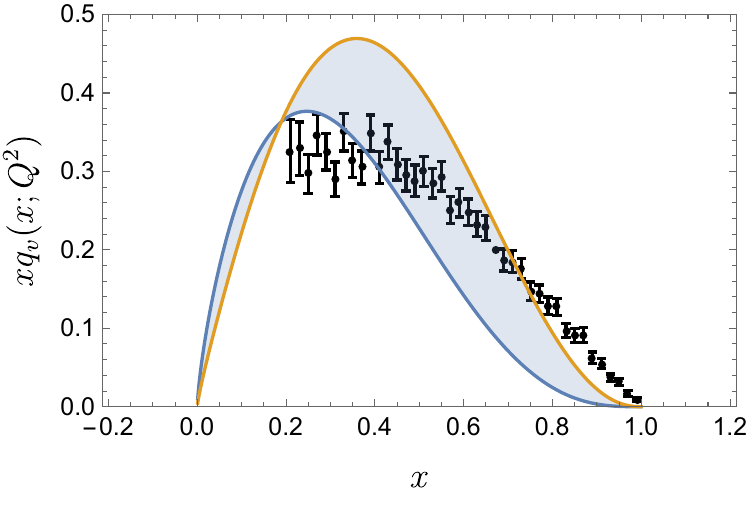} \hskip 0.2cm
\includegraphics[scale=0.6]{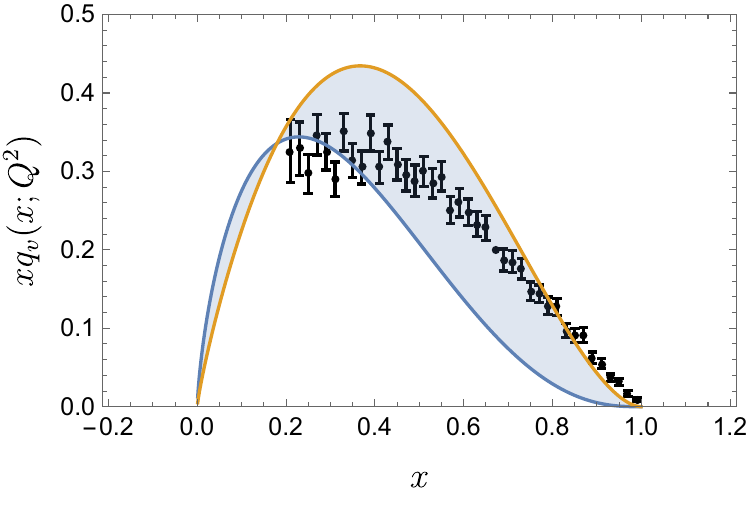}
\caption{\footnotesize  {
The pion PDF evaluated at LO to the final scale  $\mu^2 =  27$
GeV$^2$. The band stands for the uncertainty related to initial scale 
$0.07 \leq \mu_0^2 \leq 0.15$ GeV$^2$. Data are from Ref. \cite{Conway:1989fs}.
Left panel for the GSWL1 reparametrization. Right panel for the GSWL2 parametrization.
 }}
\label{fig:PDF}
\end{figure*}

As already discussed, the initial scale is fixed by fitting 
 the data of Ref. \cite{Conway:1989fs} obtained at the final scale $\mu^2 =  27$
GeV$^2$. We consider here leading order (LO)  pQCD evolution of the pion PDF. In Fig. \ref{fig:PDF} one can see that
both parametrizations are almost able to describe the data in the valence region
for small initial scales, as expected for a LO calculation. However,
in order to describe the data, higher values of $\mu_0^2$ are needed.
In Fig. \ref{fig:PDF} we display the results for
 $0.07 \leq \mu_0^2 \leq 0.15$ GeV$^2$.
One should notice that the allowed
range of initial scales here considered are similar to those usually 
adopted by constituent quark models, namely when only the constituent quarks carry all 
the momentum of the parent hadron. Thus, to conclude, we can safely state that the GSW model, including the 
longitudinal dynamics, is a very promising model to investigate the pion structure functions.

\section{Conclusion}

The GSW model was born to describe the scalar and tensor glueballs in $AdS_5$
space. It is characterized  by a warped metric and the description of these
glueballs as gravitons in five dimensions~
\cite{Rinaldi:2017wdn,Rinaldi:2018yhf,Rinaldi:2020ssz}. The model has been
 improved and extended to describe all mesons and
glueballs~\cite{Rinaldi:2021dxh}. In the latter study we appreciated that the
ground state of the pion was very peculiar and required a special dilaton in order to achieve its low mass. This feature is associated with the realization of chiral symmetry  in QCD and certainly our previous procedure, describing this mechanism with a sophisticated dilaton,  produced a pion spectrum which had many intermediate states~\cite{Rinaldi:2021dxh}, which in some sense was contradicting the data, although one must be aware that the data have a complicated structure which could hide these intermediate states. { Moreover, the wave function we could derive from its mode function is not able to provide many observables with precision.}

In this work we have adopted a different approach closer to the realization of chiral symmetry in QCD.  For that purpose we have  introduced via dilatons the appropriate scales to generate a massless pion. Thereafter  we break explicitly chiral symmetry by introducing massive quarks via longitudinal dynamics, as proposed in Refs.~\cite{Li:2021jqb,deTeramond:2021yyi}.  With this modification we  successfully  reproduced the spectrum with the addition of a well known longitudinal potential with only two additional parameters, a constituent quark mass $m_q$ and a correction to the energy scale $\gamma_\pi$ to generate the adequate mass gaps of the excitations.

Having this model for the pion we have proceeded to calculate many pion observables with only these two parameters. We discussed two different characterizations which have almost the same quark mass but quite different $\gamma_\pi$.  The relation between the observables and the parametrizations is highly non linear. The studied observables comprise low energy as well as high energy properties. We have implemented evolution, using the model calculations at a low momentum scale, in order to be able to compare with perturbative QCD results.

The main conclusion of our investigation is that, although each parametrization does better in some observables than the other, on the overall, both parametrizations lead to a good qualitative description of all the observables.

\section*{Acknowledgements}
The work was supported in part  by $i)$ Ministerio de Ciencia e Innovaci\'on and Agencia Estatal de Investigaci\'on;
of Spain MCIN/AEI/10.13039/501100011033 and European Regional Development Fund Grant No. PID2019-105439 GB-C21
$ii)$ the European Union Horizon 2020 research and innovation programme under
grant agreement STRONG - 2020 - No 824093, and $iii)$
 the European Research Council under
the European Union as Horizon 2020 research and innovation program (Grant agreement No. 804480).

\begin{appendices}

\section{}
\label{AppA}
Here we explicitly show how to derive the relations 
between th moments of the DA, see Eq. (\ref{Eq:mome0}) and 
the coefficients of the ERBL evolution expansion, see 
Eqs. (\ref{Eq:Evo1}-\ref{Eq:Evo2}). The result of this procedure
is described in
 Eq. (\ref{Eq:mome}). 
We remark that since we are interested in moments at high 
energy scales, the soft part of the DA can be neglected, 
see Eq. (\ref{Eq:DA_asy}).
In order to fulfill this properties, already 
discussed in { Ref. \cite{Choi:2007yu}}, we consider that,

\begin{align}
   \int_0^1 dx~ \phi(x;Q^2) &=\sum_n a_n(Q^2)
  \int_0^1    x (1-x) C_n^{3/2}(2x-1)
\\
\nonumber
&=\sum_n a_n(Q^2)
\dfrac{1}{6}\delta_{n 0} = \dfrac{a_0(\mu_0^2)}{6}
\\
\nonumber
&= \int_0^1 dx~ \phi^{asi}(x;\mu_0^2) = 
\dfrac{f_\pi}{2 \sqrt{3}}~.
\end{align}

Then for $\langle z^2 \rangle$ we have:

\begin{align}
   \int_0^1 dx~ (2x-1)^2 \phi(x;Q^2) &=\sum_n  a_n(Q^2)
  \int_0^1 (2x-1)^2   x (1-x) C_n^{3/2}(2x-1)
\\
\nonumber
&=\sum_n a_n(Q^2) \left[\dfrac{2}{35}\delta_{n 2}+
\dfrac{1}{30}\delta_{n 0}     \right]
\\
\nonumber
&=    \dfrac{2}{35} a_2(Q^2)+\dfrac{1}{30} a_0(Q^2)~.
\end{align}
Recursively one would get:

\begin{align}
 \dfrac{\displaystyle  \int_0^1 dx~ (2x-1)^{2l} \phi(x;Q^2)}{
\displaystyle   
\int_0^1 dx~\phi(x;Q^2) } =6 \sum_n^{2l} 
 \dfrac{a_n(Q^2)}{a_0(Q^2)}
  \int_0^1 (2x-1)^{2l}   x (1-x) C_n^{3/2}(2x-1)~.
\end{align}

\end{appendices}

\begin{appendices}
\section{}
\label{AppB}
In this section we show the the kernel of the transition form factor 
for two virtual photon produced of momenta $q_1^2 =-Q_1^2$ and 
$q_2^2 =-Q_2^2$, respectively. In particular we consider the NLO calculation
discussed in{   Ref. \cite{Braaten:1982yp} and used in Refs. \cite{Mondal:2021czk,Choi:2019wqx}}.
To this aim let us define $Q^2=Q_1^2+Q_2^2$, $w = Q_1^2/Q^2$ and 
$z=(1-x)w+x(1-w)x$. The kernel $\bar T^{NLO}_H$ is obtained from:

\begin{align}
      t(x,w) &= \left[\dfrac{w-x}{2x-1}- \left(\dfrac{z}{2w-1}
      \right)^2 \right] \left[ \dfrac{L_1 L_3}{x}+
 \dfrac{L_2 L_3}{1-x}-\dfrac{L_1^2}{2x}-
   \dfrac{L_2^2}{2(1-x)} \right]
\\
\nonumber
&+\dfrac{1-z}{2(2w-1)}(L_1-L_2)(2L_3-L_1-L_2)+ \dfrac{3}{2}(L_3-3)
\\
\nonumber
&-\left[\dfrac{3}{2}\dfrac{w-x}{2w-1}-
\left( \dfrac{z}{2w-1}  \right)^2  \right] \left[  
 \dfrac{L_1}{x}+\dfrac{L_2}{1-x} \right]-\dfrac{1}{2} 
\dfrac{3-2z}{2w-1}(L_1-L_2)~,
\end{align}
      where:

\begin{align}
      L_1 &= \log\left(\dfrac{z}{w} \right)
\\
L_2 &= \log\left(\dfrac{z}{1-w} \right)
\\
L_3 &= \log\left(z \right)~.
\end{align}
Then one can build 

\begin{align}
      T(Q,w,x) = \dfrac{1}{Q^2} \dfrac{}{(1-x)w+x(1-w)}
\left[1+ C_F \dfrac{\alpha_s(Q^2)}{2 \pi} t(x,w) \right]
\end{align}
from which the NLO kernel is obtained via symmetrization:

\begin{align}
      \bar T_H^{NLO}(x;Q_1^2,Q_2^2) =\dfrac{T(Q,w,x)+T(Q,w,1-x)}{2}~. 
\end{align}

We recall that also in this case other logarithms are neglected 
by setting $Q$ equal to the regularization scale.
\end{appendices}

\bibliographystyle{unsrt}
\bibliography{GSWPRD20213.bib}

\begin{thebibliography}{10}

\bibitem{Maldacena:1997re}
Juan~Martin Maldacena.
\newblock {The Large N limit of superconformal field theories and
  supergravity}.
\newblock {\em Int. J. Theor. Phys.}, 38:1113--1133, 1999.
\newblock [Adv. Theor. Math. Phys.2,231(1998)].

\bibitem{Witten:1998zw}
Edward Witten.
\newblock {Anti-de Sitter space, thermal phase transition, and confinement in
  gauge theories}.
\newblock {\em Adv. Theor. Math. Phys.}, 2:505--532, 1998.

\bibitem{Fritzsch:1973pi}
H.~Fritzsch, Murray Gell-Mann, and H.~Leutwyler.
\newblock {Advantages of the Color Octet Gluon Picture}.
\newblock {\em Phys. Lett.}, 47B:365--368, 1973.

\bibitem{Fritzsch:1975wn}
Harald Fritzsch and Peter Minkowski.
\newblock {Heavy Elementary Fermions and Proton Stability in Unified Theories}.
\newblock {\em Phys. Lett.}, 56B:69--72, 1975.

\bibitem{Rinaldi:2017wdn}
Matteo Rinaldi and Vicente Vento.
\newblock {Scalar and Tensor Glueballs as Gravitons}.
\newblock {\em Eur. Phys. J.}, A54:151, 2018.

\bibitem{Rinaldi:2018yhf}
Matteo Rinaldi and Vicente Vento.
\newblock {Pure glueball states in a Light-Front holographic approach}.
\newblock {\em J. Phys.}, G47(5):055104, 2020.

\bibitem{Rinaldi:2020ssz}
Matteo Rinaldi and Vicente Vento.
\newblock {Scalar spectrum in a graviton soft wall model}.
\newblock {\em J. Phys.}, G47(12):125003, 2020.

\bibitem{Rinaldi:2021dxh}
Matteo Rinaldi and Vicente Vento.
\newblock {Meson and glueball spectroscopy within the graviton soft wall
  model}.
\newblock {\em Phys. Rev. D}, 104(3):034016, 2021.

\bibitem{Polchinski:2000uf}
Joseph Polchinski and Matthew~J. Strassler.
\newblock {The String dual of a confining four-dimensional gauge theory}.
\newblock 2000.

\bibitem{Brodsky:2003px}
Stanley~J. Brodsky and Guy~F. de~Teramond.
\newblock {Light-front hadron dynamics and AdS/CFT correspondence}.
\newblock {\em Phys. Lett.}, B582:211--221, 2004.

\bibitem{DaRold:2005mxj}
Leandro Da~Rold and Alex Pomarol.
\newblock {Chiral symmetry breaking from five dimensional spaces}.
\newblock {\em Nucl. Phys.}, B721:79--97, 2005.

\bibitem{Karch:2006pv}
Andreas Karch, Emanuel Katz, Dam~T. Son, and Mikhail~A. Stephanov.
\newblock {Linear confinement and AdS/QCD}.
\newblock {\em Phys. Rev.}, D74:015005, 2006.

\bibitem{Erlich:2005qh}
Joshua Erlich, Emanuel Katz, Dam~T. Son, and Mikhail~A. Stephanov.
\newblock {QCD and a holographic model of hadrons}.
\newblock {\em Phys. Rev. Lett.}, 95:261602, 2005.

\bibitem{Gherghetta:2009ac}
Tony Gherghetta, Joseph~I. Kapusta, and Thomas~M. Kelley.
\newblock {Chiral symmetry breaking in the soft-wall AdS/QCD model}.
\newblock {\em Phys. Rev.}, D79:076003, 2009.

\bibitem{Vega:2016gip}
Alfredo Vega and Paulina Cabrera.
\newblock {Family of dilatons and metrics for AdS/QCD models}.
\newblock {\em Phys. Rev.}, D93(11):114026, 2016.

\bibitem{Sakai:2004cn}
Tadakatsu Sakai and Shigeki Sugimoto.
\newblock {Low energy hadron physics in holographic QCD}.
\newblock {\em Prog. Theor. Phys.}, 113:843--882, 2005.

\bibitem{Hirn:2005nr}
Johannes Hirn and Veronica Sanz.
\newblock {Interpolating between low and high energy QCD via a 5-D Yang-Mills
  model}.
\newblock {\em JHEP}, 12:030, 2005.

\bibitem{Hirn:2005ub}
Johannes Hirn and Veronica Sanz.
\newblock {The A(5) and the pion field}.
\newblock {\em Nucl. Phys. B Proc. Suppl.}, 164:273--276, 2007.

\bibitem{tHooft:1974pnl}
Gerard 't~Hooft.
\newblock {A Two-Dimensional Model for Mesons}.
\newblock {\em Nucl. Phys. B}, 75:461--470, 1974.

\bibitem{Li:2015zda}
Yang Li, Pieter Maris, Xingbo Zhao, and James~P. Vary.
\newblock {Heavy Quarkonium in a Holographic Basis}.
\newblock {\em Phys. Lett. B}, 758:118--124, 2016.

\bibitem{Burkardt:1997de}
M.~Burkardt.
\newblock {Mesons in a collinear QCD model}.
\newblock {\em Phys. Rev. D}, 56:7105--7118, 1997.

\bibitem{Li:2021jqb}
Yang Li and James~P. Vary.
\newblock {Light-front holography with chiral symmetry breaking}.
\newblock {\em Phys. Lett. B}, 825:136860, 2022.

\bibitem{deTeramond:2021yyi}
Guy~F. de~Teramond and Stanley~J. Brodsky.
\newblock {Longitudinal dynamics and chiral symmetry breaking in holographic
  light-front QCD}.
\newblock {\em Phys. Rev. D}, 104(11):116009, 2021.

\bibitem{Brodsky:1997de}
Stanley~J. Brodsky, Hans-Christian Pauli, and Stephen~S. Pinsky.
\newblock {Quantum chromodynamics and other field theories on the light cone}.
\newblock {\em Phys. Rept.}, 301:299--486, 1998.

\bibitem{Brodsky:2006uqa}
Stanley~J. Brodsky and Guy~F. de~Teramond.
\newblock {Hadronic spectra and light-front wavefunctions in holographic QCD}.
\newblock {\em Phys. Rev. Lett.}, 96:201601, 2006.

\bibitem{Brodsky:2007hb}
Stanley~J. Brodsky and Guy~F. de~Teramond.
\newblock {Light-Front Dynamics and AdS/QCD Correspondence: The Pion Form
  Factor in the Space- and Time-Like Regions}.
\newblock {\em Phys. Rev.}, D77:056007, 2008.

\bibitem{Andreev:2006vy}
Oleg Andreev.
\newblock {1/q**2 corrections and gauge/string duality}.
\newblock {\em Phys. Rev. D}, 73:107901, 2006.

\bibitem{Capossoli:2015ywa}
Eduardo Folco~Capossoli and Henrique Boschi-Filho.
\newblock {Glueball spectra and Regge trajectories from a modified holographic
  softwall model}.
\newblock {\em Phys. Lett.}, B753:419--423, 2016.

\bibitem{FolcoCapossoli:2019imm}
Eduardo Folco~Capossoli, Miguel~Angel Martin~Contreras, Danning Li, Alfredo
  Vega, and Henrique Boschi-Filho.
\newblock {Hadronic spectra from deformed AdS backgrounds}.
\newblock {\em Chin. Phys.}, C44(6):064104, 2020.

\bibitem{MartinContreras:2021yfz}
Miguel~Angel Martin~Contreras, Eduardo Folco~Capossoli, Danning Li, Alfredo
  Vega, and Henrique Boschi-Filho.
\newblock {Pion form factor from an AdS deformed background}.
\newblock {\em Nucl. Phys. B}, 977:115726, 2022.

\bibitem{Ghoroku:2005vt}
Kazuo Ghoroku, Nobuhito Maru, Motoi Tachibana, and Masanobu Yahiro.
\newblock {Holographic model for hadrons in deformed AdS(5) background}.
\newblock {\em Phys. Lett. B}, 633:602--606, 2006.

\bibitem{Colangelo:2007pt}
P.~Colangelo, F.~De~Fazio, F.~Jugeau, and S.~Nicotri.
\newblock {On the light glueball spectrum in a holographic description of QCD}.
\newblock {\em Phys. Lett.}, B652:73--78, 2007.

\bibitem{Contreras:2018hbi}
Miguel~Angel Martin~Contreras, Alfredo Vega, and Santiago Cortes.
\newblock {Light pseudoscalar and axial spectroscopy using AdS/QCD modified
  soft wall model}.
\newblock {\em Chin. J. Phys.}, 66:715--723, 2020.

\bibitem{Tanabashi:2018oca}
M.~Tanabashi et~al.
\newblock {Review of Particle Physics}.
\newblock {\em Phys. Rev.}, D98(3):030001, 2018.

\bibitem{Zyla:2020zbs}
P.~A. Zyla et~al.
\newblock {Review of Particle Physics}.
\newblock {\em PTEP}, 2020(8):083C01, 2020.

\bibitem{Diehl:2000xz}
M.~Diehl, T.~Feldmann, R.~Jakob, and P.~Kroll.
\newblock {The overlap representation of skewed quark and gluon distributions}.
\newblock {\em Nucl. Phys. B}, 596:33--65, 2001.
\newblock [Erratum: Nucl.Phys.B 605, 647--647 (2001)].

\bibitem{Brodsky:2008pf}
Stanley~J. Brodsky and Guy~F. de~Teramond.
\newblock {Light-Front Dynamics and AdS/QCD Correspondence: Gravitational Form
  Factors of Composite Hadrons}.
\newblock {\em Phys. Rev.}, D78:025032, 2008.

\bibitem{deTeramond:2008ht}
Guy~F. de~Teramond and Stanley~J. Brodsky.
\newblock {Light-Front Holography: A First Approximation to QCD}.
\newblock {\em Phys. Rev. Lett.}, 102:081601, 2009.

\bibitem{Drell:1969km}
S.~D. Drell and Tung-Mow Yan.
\newblock {Connection of Elastic Electromagnetic Nucleon Form-Factors at Large
  Q**2 and Deep Inelastic Structure Functions Near Threshold}.
\newblock {\em Phys. Rev. Lett.}, 24:181--185, 1970.

\bibitem{Li:2022izo}
Yang Li and James~P. Vary.
\newblock {Longitudinal dynamics for mesons on the light cone}.
\newblock 2 2022.

\bibitem{Barry:2022itu}
P.~C. Barry et~al.
\newblock {Complementarity of experimental and lattice QCD data on pion parton
  distributions}.
\newblock 4 2022.

\bibitem{Polchinski:2002jw}
Joseph Polchinski and Matthew~J. Strassler.
\newblock {Deep inelastic scattering and gauge / string duality}.
\newblock {\em JHEP}, 05:012, 2003.

\bibitem{Hong:2004sa}
Sungho Hong, Sukjin Yoon, and Matthew~J. Strassler.
\newblock {On the couplings of vector mesons in AdS / QCD}.
\newblock {\em JHEP}, 04:003, 2006.

\bibitem{Erlich2:2005qh}
Joshua Erlich, Emanuel Katz, Dam~T. Son, and Mikhail~A. Stephanov.
\newblock {QCD and a holographic model of hadrons}.
\newblock {\em Phys. Rev. Lett.}, 95:261602, 2005.

\bibitem{Huang:2007fv}
Song He, Mei Huang, Qi-Shu Yan, and Yi~Yang.
\newblock {Confront Holographic QCD with Regge Trajectories}.
\newblock {\em Eur. Phys. J.}, C66:187--196, 2010.

\bibitem{Zou:2018eam}
Liping Zou and H.~G. Dosch.
\newblock {A very Practical Guide to Light Front Holographic QCD}.
\newblock 2018.

\bibitem{West:1970av}
Geoffrey~B. West.
\newblock {Phenomenological model for the electromagnetic structure of the
  proton}.
\newblock {\em Phys. Rev. Lett.}, 24:1206--1209, 1970.

\bibitem{NA7:1986vav}
S.~R. Amendolia et~al.
\newblock {A Measurement of the Space - Like Pion Electromagnetic Form-Factor}.
\newblock {\em Nucl. Phys. B}, 277:168, 1986.

\bibitem{Ackermann:1977rp}
H.~Ackermann, T.~Azemoon, W.~Gabriel, H.~D. Mertiens, H.~D. Reich, G.~Specht,
  F.~Janata, and D.~Schmidt.
\newblock {Determination of the Longitudinal and the Transverse Part in pi+
  Electroproduction}.
\newblock {\em Nucl. Phys. B}, 137:294--300, 1978.

\bibitem{Bebek:1977pe}
C.~J. Bebek et~al.
\newblock {Electroproduction of single pions at low epsilon and a measurement
  of the pion form-factor up to $q^2$ = 10-GeV$^2$}.
\newblock {\em Phys. Rev. D}, 17:1693, 1978.

\bibitem{JeffersonLabFpi:2007vir}
V.~Tadevosyan et~al.
\newblock {Determination of the pion charge form-factor for Q**2 = 0.60-GeV**2
  - 1.60-GeV**2}.
\newblock {\em Phys. Rev. C}, 75:055205, 2007.

\bibitem{Brauel:1979zk}
P.~Brauel, T.~Canzler, D.~Cords, R.~Felst, Guenter Grindhammer, M.~Helm, W.~D.
  Kollmann, H.~Krehbiel, and M.~Schadlich.
\newblock {\em Z. Phys. C}, 3:101, 1979.

\bibitem{JeffersonLabFpi-2:2006ysh}
T.~Horn et~al.
\newblock {Determination of the Charged Pion Form Factor at Q**2 = 1.60 and
  2.45-(GeV/c)**2}.
\newblock {\em Phys. Rev. Lett.}, 97:192001, 2006.

\bibitem{Amendolia:1984nz}
S.~R. Amendolia et~al.
\newblock {A Measurement of the Pion Charge Radius}.
\newblock {\em Phys. Lett. B}, 146:116--120, 1984.

\bibitem{ParticleDataGroup:2014cgo}
K.~A. Olive et~al.
\newblock {Review of Particle Physics}.
\newblock {\em Chin. Phys. C}, 38:090001, 2014.

\bibitem{Cui:2021aee}
Zhu-Fang Cui, Daniele Binosi, Craig~D. Roberts, and Sebastian~M. Schmidt.
\newblock {Pion charge radius from pion+electron elastic scattering data}.
\newblock {\em Phys. Lett. B}, 822:136631, 2021.

\bibitem{Ahmady:2018muv}
Mohammad Ahmady, Chandan Mondal, and Ruben Sandapen.
\newblock {Dynamical spin effects in the holographic light-front wavefunctions
  of light pseudoscalar mesons}.
\newblock {\em Phys. Rev. D}, 98(3):034010, 2018.

\bibitem{deTeramond:2018ecg}
Guy~F. de~Teramond, Tianbo Liu, Raza~Sabbir Sufian, Hans~Günter Dosch,
  Stanley~J. Brodsky, and Alexandre Deur.
\newblock {Universality of Generalized Parton Distributions in Light-Front
  Holographic QCD}.
\newblock {\em Phys. Rev. Lett.}, 120(18):182001, 2018.

\bibitem{Rinaldi:2020ybv}
Matteo Rinaldi.
\newblock {Double parton correlations in mesons within AdS/QCD soft-wall
  models: a first comparison with lattice data}.
\newblock {\em Eur. Phys. J. C}, 80(7):678, 2020.

\bibitem{Rinaldi:2015cya}
Matteo Rinaldi, Sergio Scopetta, Mareo Traini, and Vicente Vento.
\newblock {Double parton scattering: a study of the effective cross section
  within a Light-Front quark model}.
\newblock {\em Phys. Lett. B}, 752:40--45, 2016.

\bibitem{Bartalini:2018qje}
Paolo Bartalini and Jonathan~Richard Gaunt, editors.
\newblock {\em {Multiple Parton Interactions at the LHC}}, volume~29.
\newblock WSP, 2019.

\bibitem{Calucci:1999yz}
G.~Calucci and D.~Treleani.
\newblock {Proton structure in transverse space and the effective
  cross-section}.
\newblock {\em Phys. Rev. D}, 60:054023, 1999.

\bibitem{Rinaldi:2018slz}
Matteo Rinaldi and Federico~Alberto Ceccopieri.
\newblock {Hadronic structure from double parton scattering}.
\newblock {\em Phys. Rev. D}, 97(7):071501, 2018.

\bibitem{Diehl:2003ny}
M.~Diehl.
\newblock {Generalized parton distributions}.
\newblock {\em Phys. Rept.}, 388:41--277, 2003.

\bibitem{Rinaldi:2018zng}
Matteo Rinaldi, Sergio Scopetta, Marco Traini, and Vicente Vento.
\newblock {A model calculation of double parton distribution functions of the
  pion}.
\newblock {\em Eur. Phys. J. C}, 78(9):781, 2018.

\bibitem{Bali:2018nde}
Gunnar~S. Bali, Peter~C. Bruns, Luca Castagnini, Markus Diehl, Jonathan~R.
  Gaunt, Benjamin Gl\"a\ss{}le, Andreas Sch\"afer, Andr\'e Sternbeck, and
  Christian Zimmermann.
\newblock {Two-current correlations in the pion on the lattice}.
\newblock {\em JHEP}, 12:061, 2018.

\bibitem{Courtoy:2020tkd}
Aurore Courtoy, Santiago Noguera, and Sergio Scopetta.
\newblock {Two-current correlations in the pion in the Nambu and Jona-Lasinio
  model}.
\newblock {\em Eur. Phys. J. C}, 80(10):909, 2020.

\bibitem{Courtoy:2019cxq}
Aurore Courtoy, Santiago Noguera, and Sergio Scopetta.
\newblock {Double parton distributions in the pion in the
  Nambu\textendash{}Jona-Lasinio model}.
\newblock {\em JHEP}, 12:045, 2019.

\bibitem{Broniowski:2019rmu}
Wojciech Broniowski and Enrique Ruiz~Arriola.
\newblock {Double parton distribution of valence quarks in the pion in chiral
  quark models}.
\newblock {\em Phys. Rev. D}, 101(1):014019, 2020.

\bibitem{Lepage:1980fj}
G.~Peter Lepage and Stanley~J. Brodsky.
\newblock {Exclusive Processes in Perturbative Quantum Chromodynamics}.
\newblock {\em Phys. Rev. D}, 22:2157, 1980.

\bibitem{Mondal:2021czk}
Chandan Mondal, Sreeraj Nair, Shaoyang Jia, Xingbo Zhao, and James~P. Vary.
\newblock {Pion to photon transition form factors with basis light-front
  quantization}.
\newblock {\em Phys. Rev. D}, 104(9):094034, 2021.

\bibitem{Brodsky:1984xk}
Stanley~J. Brodsky, P.~Damgaard, Y.~Frishman, and G.~Peter Lepage.
\newblock {CONFORMAL SYMMETRY: EXCLUSIVE PROCESSES BEYOND LEADING ORDER}.
\newblock {\em Phys. Rev. D}, 33:1881, 1986.

\bibitem{Radyushkin:1977gp}
A.~V. Radyushkin.
\newblock {Deep Elastic Processes of Composite Particles in Field Theory and
  Asymptotic Freedom}.
\newblock 6 1977.

\bibitem{Brodsky:2011yv}
Stanley~J. Brodsky, Fu-Guang Cao, and Guy~F. de~Teramond.
\newblock {Evolved QCD predictions for the meson-photon transition form
  factors}.
\newblock {\em Phys. Rev. D}, 84:033001, 2011.

\bibitem{Efremov:1979qk}
A.~V. Efremov and A.~V. Radyushkin.
\newblock {Factorization and Asymptotical Behavior of Pion Form-Factor in QCD}.
\newblock {\em Phys. Lett. B}, 94:245--250, 1980.

\bibitem{E791:2000xcx}
E.~M. Aitala et~al.
\newblock {Direct measurement of the pion valence quark momentum distribution,
  the pion light cone wave function squared}.
\newblock {\em Phys. Rev. Lett.}, 86:4768--4772, 2001.

\bibitem{Choi:2007yu}
Ho-Meoyng Choi and Chueng-Ryong Ji.
\newblock {Distribution amplitudes and decay constants for (pi, K, rho, K*)
  mesons in light-front quark model}.
\newblock {\em Phys. Rev. D}, 75:034019, 2007.

\bibitem{Arthur:2010xf}
R.~Arthur, P.~A. Boyle, D.~Brommel, M.~A. Donnellan, J.~M. Flynn, A.~Juttner,
  T.~D. Rae, and C.~T.~C. Sachrajda.
\newblock {Lattice Results for Low Moments of Light Meson Distribution
  Amplitudes}.
\newblock {\em Phys. Rev. D}, 83:074505, 2011.

\bibitem{Braun:2015axa}
V.~M. Braun, S.~Collins, M.~G\"ockeler, P.~P\'erez-Rubio, A.~Sch\"afer, R.~W.
  Schiel, and A.~Sternbeck.
\newblock {Second Moment of the Pion Light-cone Distribution Amplitude from
  Lattice QCD}.
\newblock {\em Phys. Rev. D}, 92(1):014504, 2015.

\bibitem{Braun:2006dg}
V.~M. Braun et~al.
\newblock {Moments of pseudoscalar meson distribution amplitudes from the
  lattice}.
\newblock {\em Phys. Rev. D}, 74:074501, 2006.

\bibitem{Bali:2017ude}
G.~S. Bali, V.~M. Braun, M.~G\"ockeler, M.~Gruber, F.~Hutzler, P.~Korcyl,
  B.~Lang, and A.~Sch\"afer.
\newblock {Second moment of the pion distribution amplitude with the momentum
  smearing technique}.
\newblock {\em Phys. Lett. B}, 774:91--97, 2017.

\bibitem{RQCD:2019osh}
Gunnar~S. Bali, Vladimir~M. Braun, Simon B\"urger, Meinulf G\"ockeler, Michael
  Gruber, Fabian Hutzler, Piotr Korcyl, Andreas Sch\"afer, Andr\'e Sternbeck,
  and Philipp Wein.
\newblock {Light-cone distribution amplitudes of pseudoscalar mesons from
  lattice QCD}.
\newblock {\em JHEP}, 08:065, 2019.
\newblock [Addendum: JHEP 11, 037 (2020)].

\bibitem{Zhang:2020gaj}
Rui Zhang, Carson Honkala, Huey-Wen Lin, and Jiunn-Wei Chen.
\newblock {Pion and kaon distribution amplitudes in the continuum limit}.
\newblock {\em Phys. Rev. D}, 102(9):094519, 2020.

\bibitem{Belle:2012wwz}
S.~Uehara et~al.
\newblock {Measurement of $\gamma \gamma^* \to \pi^0$ transition form factor at
  Belle}.
\newblock {\em Phys. Rev. D}, 86:092007, 2012.

\bibitem{BaBar:2009rrj}
Bernard Aubert et~al.
\newblock {Measurement of the gamma gamma* ---\ensuremath{>} pi0 transition
  form factor}.
\newblock {\em Phys. Rev. D}, 80:052002, 2009.

\bibitem{CLEO:1997fho}
J.~Gronberg et~al.
\newblock {Measurements of the meson - photon transition form-factors of light
  pseudoscalar mesons at large momentum transfer}.
\newblock {\em Phys. Rev. D}, 57:33--54, 1998.

\bibitem{CELLO:1990klc}
H.~J. Behrend et~al.
\newblock {A Measurement of the pi0, eta and eta-prime electromagnetic
  form-factors}.
\newblock {\em Z. Phys. C}, 49:401--410, 1991.

\bibitem{Cao:1996di}
Fu-Guang Cao, Tao Huang, and Bo-Qiang Ma.
\newblock {The Perturbative pion - photon transition form-factors with
  transverse momentum corrections}.
\newblock {\em Phys. Rev. D}, 53:6582--6585, 1996.

\bibitem{Musatov:1997pu}
I.~V. Musatov and A.~V. Radyushkin.
\newblock {Transverse momentum and Sudakov effects in exclusive QCD processes:
  Gamma* gamma pi0 form-factor}.
\newblock {\em Phys. Rev. D}, 56:2713--2735, 1997.

\bibitem{Brodsky:1981rp}
Stanley~J. Brodsky and G.~Peter Lepage.
\newblock {Large Angle Two Photon Exclusive Channels in Quantum
  Chromodynamics}.
\newblock {\em Phys. Rev. D}, 24:1808, 1981.

\bibitem{Braaten:1982yp}
Eric Braaten.
\newblock {QCD CORRECTIONS TO MESON - PHOTON TRANSITION FORM-FACTORS}.
\newblock {\em Phys. Rev. D}, 28:524, 1983.

\bibitem{Chang:2020kjj}
Lei Chang, Kh\'epani Raya, and Xiaobin Wang.
\newblock {Pion Parton Distribution Function in Light-Front Holographic QCD}.
\newblock {\em Chin. Phys. C}, 44(11):114105, 2020.

\bibitem{Lan:2019rba}
Jiangshan Lan, Chandan Mondal, Shaoyang Jia, Xingbo Zhao, and James~P. Vary.
\newblock {Pion and kaon parton distribution functions from basis light front
  quantization and QCD evolution}.
\newblock {\em Phys. Rev. D}, 101(3):034024, 2020.

\bibitem{dePaula:2022pcb}
W.~de~Paula, E.~Ydrefors, J.~H. Nogueira~Alvarenga, T.~Frederico, and
  G.~Salm\`e.
\newblock {Parton distribution function in a pion with Minkowskian dynamics}.
\newblock {\em Phys. Rev. D}, 105(7):L071505, 2022.

\bibitem{Noguera:2015iia}
Santiago Noguera and Sergio Scopetta.
\newblock {Pion transverse momentum dependent parton distributions in the Nambu
  and Jona-Lasinio model}.
\newblock {\em JHEP}, 11:102, 2015.

\bibitem{Courtoy:2020fex}
Aurore Courtoy and Pavel~M. Nadolsky.
\newblock {Testing momentum dependence of the nonperturbative hadron structure
  in a global QCD analysis}.
\newblock {\em Phys. Rev. D}, 103(5):054029, 2021.

\bibitem{Conway:1989fs}
J.~S. Conway et~al.
\newblock {Experimental Study of Muon Pairs Produced by 252-GeV Pions on
  Tungsten}.
\newblock {\em Phys. Rev. D}, 39:92--122, 1989.

\bibitem{Choi:2019wqx}
Ho-Meoyng Choi, Hui-Young Ryu, and Chueng-Ryong Ji.
\newblock {Doubly virtual $(\pi^0,\eta,\eta')\to\gamma^*\gamma^*$ transition
  form factors in the light-front quark model}.
\newblock {\em Phys. Rev. D}, 99(7):076012, 2019.

\end{thebibliography}

\end{document}